\documentclass[twocolumn,superscriptaddress,aps,epsf,pra,showpacs]{revtex4}
\newcommand{\lb}{\left [ }
\newcommand{\rb}{\right ] }

\newcommand{\ket}[1]{\left |#1\right >}
\newcommand{\opsandwich}[3]{\left < #1\left|#2\right|#3\right >}

\newcommand{\expec}[1]{\left < #1\right >}

\usepackage{amsmath}
\usepackage{amsfonts}
\usepackage{graphicx}

\usepackage{hyperref}

\usepackage[usenames]{color}

\renewcommand{\v}[1]{\boldsymbol{#1}}

\newcommand{\lp}{\left ( }
\newcommand{\rp}{\right ) }

\newcommand{\hc}{\text{H.c.}}

\newcommand{\mc}{\mathcal}

\newcommand{\beq}{\begin{eqnarray*}}
\newcommand{\eeq}{\end{eqnarray*}}

\newcommand{\be}{\begin{eqnarray}}
\newcommand{\ee}{\end{eqnarray}}

\def\lsim{\mathrel{\rlap{\lower4pt\hbox{\hskip1pt$\sim$}}
    \raise1pt\hbox{$<$}}}                

\def\gsim{\mathrel{\rlap{\lower4pt\hbox{\hskip1pt$\sim$}}
    \raise1pt\hbox{$>$}}}                

\begin{document}

\title{Spectroscopy of dipolar fermions in 2D pancakes and 3D lattices}
\author{Kaden R. A. Hazzard} \email{kaden.hazzard@colorado.edu}
\affiliation{JILA and Department of Physics, University of Colorado, Boulder, and NIST, Boulder, Colorado 80309-0440, USA}
\author{Alexey V. Gorshkov}
\affiliation{Institute for Quantum Information, California Institute of Technology, Pasadena, California 91125, USA}
\author{Ana Maria Rey}
\affiliation{JILA and Department of Physics, University of Colorado, Boulder, and NIST, Boulder, Colorado 80309-0440, USA}

\begin{abstract}
Motivated by ongoing measurements at JILA, we calculate the recoil-free  spectra of dipolar interacting fermions, for example ultracold heteronuclear molecules, in a one-dimensional lattice of two-dimensional pancakes, spectroscopically probing transitions between different internal (e.g., rotational) states.  We additionally incorporate $p$-wave interactions and losses, which are important for reactive molecules such as KRb.  Moreover, we consider other sources of spectral broadening: interaction-induced quasiparticle lifetimes and the different polarizabilities of the different rotational states used for the spectroscopy.  Although our main focus is molecules, some of the calculations are also useful for optical lattice atomic clocks. For example, understanding the $p$-wave shifts between identical fermions and small dipolar interactions coming from the excited clock state are necessary to reach future precision goals.  Finally, we consider the spectra in a deep 3D lattice and show how they give a great deal of information about static correlation functions, including \textit{all} the moments of the density correlations between nearby sites. The range of correlations measurable depends on spectroscopic resolution and the dipole moment.
\end{abstract}


\maketitle

\section{Introduction}

Recoil-free spectroscopy is a powerful, pervasive probe of ultracold atomic systems that includes radio-frequency and microwave spectroscopy, as well as Doppler-free optical or Raman spectroscopy.  Interatomic interaction effects on the spectra give information about the phase diagram and behavior of many-body systems.  For example, these effects on recoil-free spectra were used to first detect Bose-Einstein condensation in dilute spin-polarized atomic hydrogen~\cite{fried:hydrogen}, locate and probe the Mott insulator/superfluid quantum phase transition of bosons in optical lattices~\cite{campbell_imagingmott_2006-2,ohashi_itinerant-localized_2006,hazzard_hyperfine_2007,hazzard_many-body_2009,sun_probing_2009}, and study Cooper pair binding~\cite{schunck:determination_2008,schirotzek:determination_2008}, polaron quasiparticle residue~\cite{schirotzek:observation_2009}, and pseudogap behavior of ultracold fermions across the BEC/BCS crossover~\cite{stewart:using_2008,gaebler:observation_2010,perali:evolution_2011}.
In precision measurements, e.g. atomic clocks, understanding the effects of interactions is crucial because they can limit~\cite{lemke:spin_2009} or enhance~\cite{swallows:suppression_2011} the achieved accuracy.

Here we calculate the recoil-free spectra of dipolar molecules in a one-dimensional lattice of two-dimensional ``pancakes" and in a three-dimensional lattice. Such systems have been realized at JILA~\cite{miranda:controlling_2011,ospelkaus:quantum-state_2010,ni:a_2008,ni:dipolar_2010,ospelkaus:ultracold_2009}.
As we demonstrate, recoil-free spectroscopy gives information complementary to density profile or time-of-flight measurements.  Additional probes are especially important for molecules, where direct imaging is extremely difficult due to the absence of cycling transitions~\cite{wang_direct_2010}.  To be concrete, we consider driving the transition between two rotational states of the molecule, but the same calculations apply to transitions between vibrational, hyperfine, and electronic states when the transitions are measured with techniques imparting negligible recoil momentum.

There is great excitement about ultracold molecules: they offer additional internal degrees of freedom (rotational, vibrational) compared to atoms, and the long-range character of their interactions can lead to interesting many-body physics~\cite{lahaye_physics_2009,pupillo:condensed_2008,baranov:theoretical_2008,carr:cold_2009}. Examples are quantum liquid crystals~\cite{lin:liquid_2010,quintanilla:metanematic_2009,sun:spontaneous_2010}, Wigner crystals~\cite{cheng:static_2009,baranov:wigner_2008}, exotic superfluids~\cite{baranov:bilayer_2010,bruun:quantum_2010,wang:quantum_2006}, supersolids~\cite{trefzger:ultracold_2011},
topological phases~\cite{osterloh:strongly_2007,qiu:quantum_2011,cheng:excitation_2008,baranov:fractional_2005},
and the physics resulting from quantum phase transitions among this plethora of phases~\cite{hazzard:techniques_quantumcriticality,zhou:signature_2010}.  We expect recoil-free spectroscopy to play a major role in observing and characterizing these systems.

The outline of our paper is as follows. Sec.~\ref{sec:general-sumrule} gives general expressions, valid even for strongly correlated systems, relating the average spectral shifts to simple static correlation functions.  Sec.~\ref{sec:weakly-interacting-sumrule} evaluates these expressions explicitly for weakly interacting homogeneous systems. Sec.~\ref{sec:sumrule-lda} applies these expressions to obtain the trapped system's spectra, while Sec.~\ref{sec:example-spec} shows example results. These calculations are directly relevant for the ongoing experiments at JILA~\cite{miranda:controlling_2011,ospelkaus:quantum-state_2010,ni:a_2008,ni:dipolar_2010,ospelkaus:ultracold_2009} and should be useful in a number of related ongoing efforts to make ultracold molecules, e.g.~\cite{lercher:production_2011}.

We quantitatively calculate the spectral shifts and, in Sec.~\ref{sec:collisional-broadening}, lineshapes due to the dipolar interactions, $p$-wave interactions -- including $p$-wave losses that are important for reactive molecules such as KRb -- and the differing polarizabilities of the different molecular rotational states that may be used in the spectroscopy.  We also qualitatively consider and bound other effects, such as higher lattice band occupation and the validity of treating the atoms in the cold collision regime, i.e. treating interactions via a pseudopotential.

To briefly motivate the utility of these spectra, we point out that measuring them in the dilute thermal gas allows one to ensure that these various non-trivial contributions --- many of which are just beginning to be explored in the context of cold atoms  --- behave  as expected.  Furthermore, deviations from the simple dilute thermal gas values allow one to diagnose more interesting behavior as the temperature is lowered, the simplest example probably being the degree of quantum degeneracy of the gas.  The onset of strongly correlated quantum phases will also have signatures.

As another application of recoil-free spectra to dipolar interacting systems, Sec.~\ref{sec:3d-lattice} considers a deep 3D lattice.  We show that one can measure the joint probability distribution $P(1,n_1,n_2,\ldots)$ of a particle occupying a site, while having  $n_1$ molecules in the nearest neighbor sites, $n_2$ in the next-nearest neighbor sites, and so on.  This includes \textit{all} moments $\expec{n_1^{\alpha_1}n_2^{\alpha_2}\ldots}$ for arbitrary $\alpha_j$. The spatial range of correlations that can be measured is set by the spectral resolution and interaction strength, and is typically a few sites for the molecules presently considered.  This provides a powerful way to characterize the system, measuring a much more complete set of correlations and with greater spatial range  than other techniques, such as modulation spectroscopy~\cite{greif:probing_2011}.

While our quantitative estimates of parameters are for dipolar molecules, many of the calculations are also relevant for Rydberg atoms and alkaline earth atomic clock experiments.  As one example, Rydberg atoms have long range dipolar interactions.  As another,
$p$-wave interactions can be important in optical atomic clocks, as recently revealed by
Ramsey spectroscopy of Yb lattice clocks with the atoms trapped in 1D arrays of 2D pancakes or 2D arrays of 1D tubes~\cite{ludlow:pc}.  
Moreover,
as experiments in these clocks  expand their capabilities to approach expected $10^{-18}$ accuracy and mHz frequency resolution, small dipolar interactions between the atoms become relevant.  We expect our calculations may provide insight into these experiments.
However, we mention that  while our theory assumes that few atoms are transferred to the excited state (i.e., linear response), many clock interrogation techniques transfer a substantial fraction of the atoms~\cite{rey:many-body_2009,yu:clock_2010}.


\section{Theoretical and experimental background}

\begin{figure}[hbtp]
\setlength{\unitlength}{1.0in}
\includegraphics[width=3.3in,angle=0]{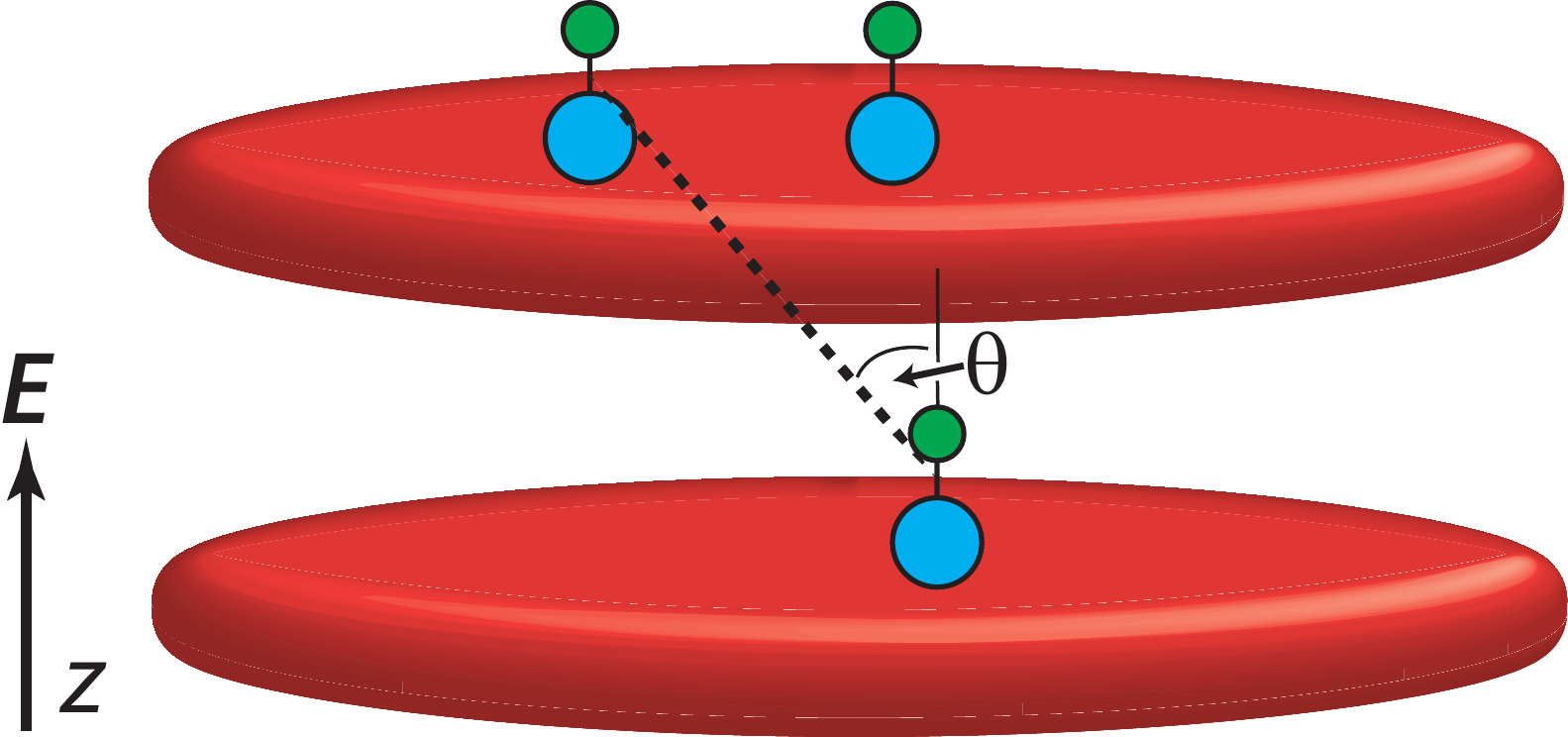}
\caption{Schematic of the experimental geometry considered.  Heteronuclear dipolar molecules are trapped in a one-dimensional lattice of two-dimensional pancakes, with the lattice periodic in the $z$ direction.  The angle $\theta$ is the angle between the inter-molecule separation $\v{r}-\v{r'}$ and the molecular dipole (applied electric field axis), here taken to be along the $z$-axis.
 \label{fig:geometry-schematic}}
\end{figure}

We are interested in heteronuclear molecules in a one dimensional optical lattice of two dimensional pancakes in the presence of an electric field.  We assume that the energy scale set by this field is large compared to the hyperfine splitting; we explain this in more detail below. Fig.~\ref{fig:geometry-schematic} illustrates this system and defines notation.  Section~\ref{sec:3d-lattice} considers the effects of adding a lattice to the transverse directions.  This system displays a hierarchy of energy scales, which dictates the physical behavior.  We begin by describing this hierarchy.

\begin{figure}[hbtp]
\setlength{\unitlength}{1.0in}
\includegraphics[width=3.3in,angle=0]{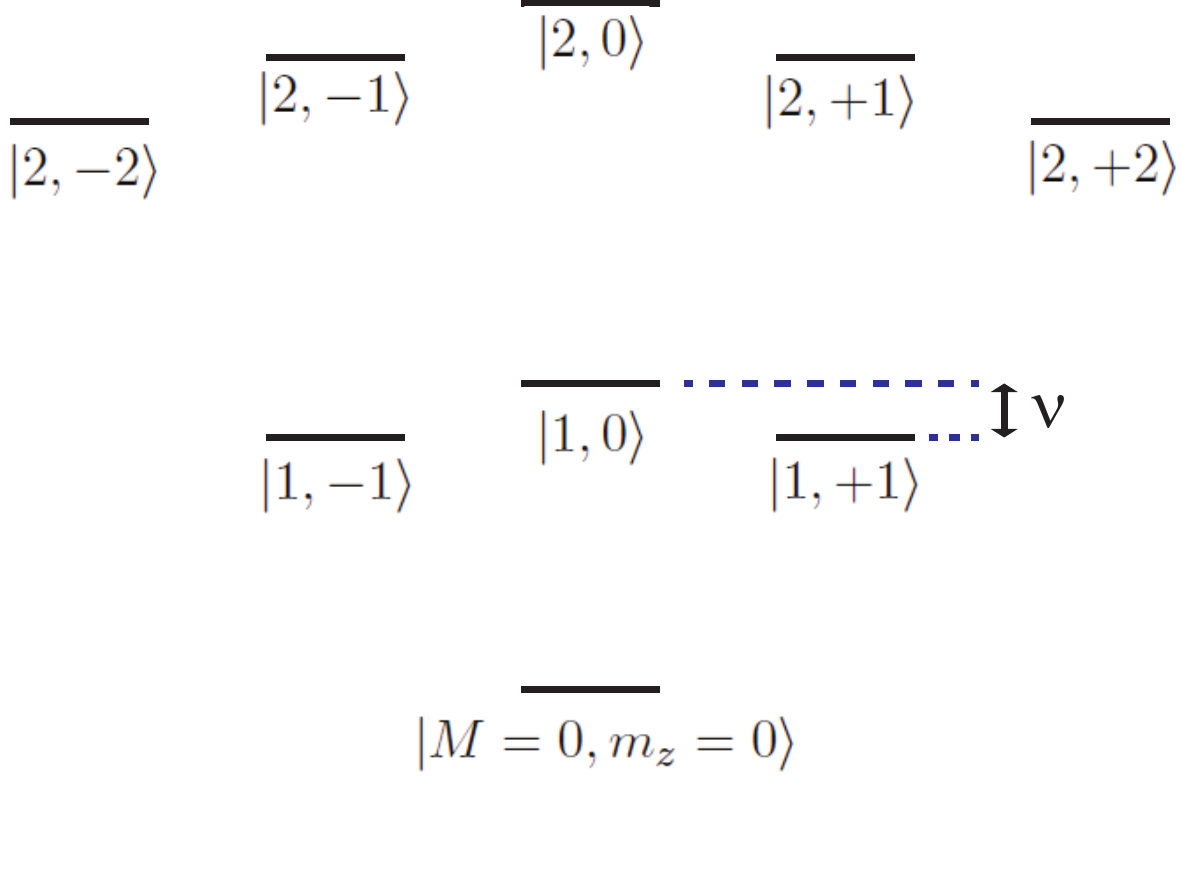}
\caption{Lowest rotational energy levels of a single molecule in an electric field.  These are described by the rigid rotor Hamiltonian in an electric field $H=BN^2-d_0 E$ with $N$ the angular momentum operator, $B$ the rotational constant, and $d_0$ the dipole moment along $z$. The label  $\ket{M, m_z}$ indicates the state that is adiabatically connected to the $E=0$ state with angular momentum $M$, and angular momentum $z$-component projection $m_z$.  However, the $E$-field mixes states with the same $m_z$ but with different $M$, so $M$ is \textit{not} a good quantum  numbers.  The energy splitting $\nu$ increases with increasing electric field and is a large energy scale even for small $E$: interactions, the trapping potential, and the kinetic energy cannot mix in states that are off-resonant by this energy scale.
\label{fig:rot-level-diagram}}
\end{figure}

Figure~\ref{fig:rot-level-diagram} shows the lowest energy rotational levels of molecule in a DC electric field.  This structure is well-described by the rigid rotor Hamiltonian in an electric field, $H=BN^2-\v{d}\cdot\v{E}$ with $N$ the angular momentum operator, $B$ the rotational constant, and $\v{d}$ the dipole moment operator.
This single-molecule physics sets the largest energy scales in the problem.  In the absence of an electric field, there is the usual rotational level structure: a ground state, a 3-fold degenerate spin-1 first excited state manifold, a 5-fold degenerate spin-2 second excited state manifold, etc.  These are labeled by their total angular momentum $M$ and $z$-axis projection $m_z$. In the presence of an electric field $\v{E}$ the rotational symmetry is broken.  Here we take $\v{E}$ to be along the $z$ direction, perpendicular to the pancakes, as is used in the ongoing KRb experiments to suppress the reactive losses~\cite{miranda:controlling_2011}.  The electric field mixes states with different $M$ but the same $m_z$.

Figure~\ref{fig:rot-level-diagram} shows the resulting level structure for a non-zero value of $E$, where we label by $\ket{M,m_z}$ the state that is adiabatically connected to the $E=0$ state with angular momentum $M$ and $z$-projection $m_z$.  An electric field induces a level splitting $\nu$ in the previously degenerate excited state manifold, and this energy scale is large compared to the other energy scales that will be of interest.  As a consequence, when we treat the interactions, lattice,  and trap, it is quantitatively accurate restrict oneself to processes occurring in the resonant subspace where we neglect transitions that require energy changes on the order of $\nu$.

We now present the Hamiltonian describing the period lattice, the harmonic confining potential, and intermolecular interactions for molecules in an electric field. We consider two molecular states, choosing $\ket{0,0}$ and $\ket{1,0}$ to be concrete, but the vast majority of our calculation is general and straightforwardly extended to arbitrary internal states.  We will denote these states as $\ket{\alpha}$ with $\alpha=1,2$ henceforth.  The hyperfine interaction energy is the next largest energy scale after the rotational and electric field, and can couple the $\alpha=1,2$ and other rotational states, shown in Fig.~\ref{fig:rot-level-diagram}. However, even a small electric field -- one that generates dipole moments more than an order of magnitude smaller than the permanent dipole moment -- suffices to generate level splittings $\sim\!\nu$  that are larger than the hyperfine coupling. Thus for even such small electric fields, if the  initial populations are confined to these two rotational states, we can neglect the other states, which are then far off-resonant compared to the hyperfine energy.

 The lattice energy scale is the largest after the rotational, electric field, and hyperfine structure of the molecules, all set by single molecule physics.  The lattice separates the energy levels into bands, and because the other energy scales -- trap, interactions, and temperature -- are small relative to the band splitting for the deep lattices of present interest, we can thus project the Hamiltonian to the lowest lattice band for states $\alpha=1,2$.
 We neglect terms that are exponentially smaller in the lattice depth (e.g. tunneling terms).
The molecular system is described by the Hamiltonian
\be
H &=& H_0 + H_{\text{sw}}+H_{\text{dipole}}+H_{\text{pw}}+H_{\text{trap}}
\ee
with the non-interacting part
\be
H_0 &=& \sum_{i,\alpha} \int\!d\v{\rho} \, \psi^\dagger_\alpha(i,\v{\rho}) \lp -\frac{\nabla^2}{2m}-\mu\rp \psi_\alpha^{\phantom{\dagger}}(i,\v{\rho}),
\ee
where $\psi_\alpha(i,\v{\rho})$ and $\psi^\dagger_\alpha(i,\v{\rho})$ are fermionic annihilation and creation operators acting on atoms in rotational state $\alpha$ at site $i$ in the lowest band at transverse two-dimensional position $\v{\rho}$.  They satisfy the anti-commutation relationships $\{\psi_{\alpha}(i,\v{\rho}),\psi_{\beta}^\dagger(j,\v{\rho'})\}=\delta_{ij}\delta_{\alpha\beta} \delta(\v{\rho}-\v{\rho'})$. Here $m$ is the molecular mass and $\mu$ is the chemical potential.

The projected $s$-wave interaction Hamiltonian is
\be
H_{\text{sw}} &=&
     \sum_{i,\alpha,\beta} \frac{g_{\alpha\beta}}{2} \int \!d\v{\rho}\, \psi_\alpha^\dagger(i,\v{\rho}) \psi_\beta^\dagger(i,\v{\rho}) \psi_\beta^{\phantom{\dagger}}(i,\v{\rho}) \psi_\alpha^{\phantom{\dagger}}(i,\v{\rho})\nonumber \\
\ee
with $g_{\alpha\beta} = (4\pi a_{\alpha\beta}/m)\int\!dz\, |w_\alpha(z)|^2|w_\beta(z)|^2 $ (we set $\hbar=k_B=1$ throughout), where $a_{\alpha\beta}$ is the $s$-wave scattering length for scattering between molecules in states $\alpha$ and $\beta$,  and $w_\alpha(z)$ is the lowest band Wannier function for particles in state $\alpha$.  This depends on $\alpha$ because the lattice depths felt by each species need not be equal since the molecular polarizabilities differ.  The interaction $g_{11}$ is undefined for identical fermions, which don't scatter in the $s$-wave channel, and so we can set it to zero, but $g_{12}$ is nonzero. When the Wannier functions for the two states are not identical, the three-dimensional $p$-wave interaction projected onto the lowest band gives a renormalization of the two-dimensional $1$-$2$ $s$-wave interaction (not included above) in addition to the contribution coming from the three-dimensional $s$-wave scattering, because the particles effectively become distinguishable. However,
the $s$-wave interaction will turn out to be irrelevant for the recoil-free spectra in the linear response regime with all molecules initially in the ground internal state, as a consequence of the Pauli exclusion principle.  Hence we can ignore the $s$-wave contact interaction.

The dipolar interaction Hamiltonian in 3D is~\cite{gorshkov:dipoles-tjvw-short,gorshkov:dipoles-tjvw-long}
\be
V^{(3d)}_{\alpha\beta}\! \! &=& \!\! \int \! d\v{r}d\v{r'}\, \frac{1-3\cos^2\theta}{4\pi\epsilon_0|\v{r}-\v{r'}|^3} \nonumber \\
&&\hspace{-0.40in}{}\times\bigg[ d_{12} d_{21}\varphi_1^\dagger(\v{r})\varphi^\dagger_2(\v{r'})
\varphi_1^{\phantom{\dagger}}(\v{r'})\varphi_2^{\phantom{\dagger}}(\v{r}) + \hc \nonumber \\
    &&\hspace{-0.28in}{}+ \sum_{mm'} d_{mm} d_{m'm'} \varphi^\dagger_m(\v{r})\varphi^\dagger_{m'}(\v{r'})
    \varphi_{m'}^{\phantom{\dagger}}(\v{r'})\varphi_{m}^{\phantom{\dagger}}(\v{r}) \bigg]\!,
\ee
where $\theta$ is the angle between the dipole orientation direction and $\v{r}-\v{r'}$, $d_{mn}=\opsandwich{m}{d^{(0)}}{n}$ where $d^{(0)}$ is the $m_z=0$ spherical component of the dipole operator in the basis aligned with the electric field, and $\varphi_\alpha(\v{r})$ and $\varphi_\beta^\dagger(\v{r})$ are fermionic annihilation and creation operators satisfying the anticommutation relations $\{\varphi_\alpha^{\phantom{\dagger}}(\v{r}),\varphi_\beta^\dagger(\v{r'})\}=
\delta_{\alpha\beta}\delta(\v{r}-\v{r'})$. The reason why only the $m=0$ spherical component of the dipole
operator is retained is that even a small electric field drives the $m=\pm1$ levels off-resonance, as
shown in Fig.~\ref{fig:rot-level-diagram}.
To be sufficiently off-resonant compared to
the other energy scales -- interactions, lattice, and hyperfine coupling (to allow one to neglect the nuclear degrees of freedom) -- requires only a small field, where the
induced dipole moment is still an order of magnitude or more smaller than the permanent dipole moment.
The same reason justifies why only the shown matrix elements of the dipole interaction are kept -- the other matrix elements leave the resonant subspace.
Note that the permanent dipole moments of these levels are non-vanishing, in contrast to the zero-field rotational levels, because the electric field induces a dipole moment to the eigenstates.

The projected dipolar interaction Hamiltonian is
\be
H_{\text{dipole}} &=& \frac{1}{2}\sum_{ij,\alpha\beta} \int \!d\v{\rho}d\v{\rho'} V_{\alpha\beta}(i-j,\v{\rho}-\v{\rho'}) \nonumber \\
    &&\hspace{0.08in}{}\times\psi_\alpha^\dagger(i,\v{\rho}) \psi_\beta^\dagger(j,\v{\rho'}) \psi_\beta^{\phantom{\dagger}}(j,\v{\rho'}) \psi_\alpha^{\phantom{\dagger}}(i,\v{\rho}) \nonumber \\
    &&\hspace{-0.45in}{}+\frac{1}{2}\sum_{ij} \int \!d\v{\rho}d\v{\rho'} V_{\text{sf}}(i-j,\v{\rho}-\v{\rho'}) \nonumber \\
    &&\hspace{-0.28in}{}\times\lp \psi_1^\dagger(i,\v{\rho})
    \psi_2^\dagger(j,\v{\rho'})\psi_1^{\phantom{\dagger}}(j,\v{\rho'}) \psi_2^{\phantom{\dagger}}(i,\v{\rho})+ \hc \rp\label{eq:proj-dipole-1}
\ee
with
\be
V_{\alpha\beta}(i-j,\v{\rho}-\v{\rho'})\!\! &=& \!\!\int \! dz dz' \frac{d_{\alpha\alpha}d_{\beta\beta}(1-3\cos^2\theta)}{4\pi\epsilon_0\!\lb \lp \v{\rho}-\v{\rho'}\rp^2 \!+ \!\lp z-z'\rp^2\rb^{3/2}} \nonumber \\
    &&{}\times |w_\alpha(z-z_i)|^2 |w_{\beta}(z'-z_j)|^2
\ee
and
\be
V_{\text{sf}} (i-j,\v{\rho}-\v{\rho'}) \!\! &=& \!\! \int \! dz dz' \frac{d_{12}d_{21}(1-3\cos^2\theta)}{4\pi \epsilon_0 \lb (\v{\rho}-\v{\rho'})^2+(z-z')^2 \rb^{3/2}} \nonumber \\
&&\hspace{-0.75in}{}\times w_1^*(z-z_i) w_2^{\phantom{*}}(z-z_i)
w_2^*(z'-z_j) w_1^{\phantom{*}}(z'-z_j).
\ee
  We refer to the former term as the ``direct" term and the second as the ``spin flip" term.
We approximate  $V_{\alpha\beta}(j,\v{\rho})$ by
\be
V_{\alpha\beta}(j,\v{\rho}) &\approx& \gamma_{\alpha\beta} {\mc U}(j,\v{\rho})
\ee
and $V_{\text{sf}}$ by
\be
V_{\text{sf}}(j,\v{\rho}) &\approx&  \eta  {\mc U}(j,\v{\rho}),
\ee
with
\be
\mc U(j,\v{\rho}) &=& \delta_{0j}\frac{1}{\rho^3+A_{\alpha\beta}} \nonumber \\
    &&\hspace{-0.5in}{}+(1-\delta_{0j}) \frac{ 1 - 3 \lp\frac{(jd_l)^2}{\rho^2+(jd_l)^2}\rp}{[\rho^2+(jd_l)^2]^{3/2}}
\ee
where $\gamma_{\alpha\beta}=d_{\alpha\alpha}  d_{\beta\beta}/(4\pi \epsilon_0)$, $\eta = \left|\int \! dz w_1^*(z) w_2(z) \right|^2 d_{12}^2/(4\pi \epsilon_0)$,  $d_l$ is the lattice spacing,
and $A_{\alpha\beta} \sim \ell^3$, with $\ell$ the width of the Wannier functions, governs the short range physics.
More details on this approximation follow shortly.
In principle, we should let $A_{\alpha \beta}$ be different for the $V_{\text{sf}}$
and $V_{\alpha\beta}$ terms, but we will argue that we can set $A_{\alpha\beta}=0$ and maintain
quantitative accuracy in the calculated spectra.

We also note that the factor of $F=\left|\int \! dz w_1^*(z) w_2(z) \right|^2 $ appearing in $\eta$ is almost always nearly unity in a deep lattice.  We denote the polarizability of state $2$ by $(1+\xi)$ times the polarizability of state $1$.  In a deep lattice $w_\alpha(z)$ can be approximated by a Gaussian.  Then one finds $F = \sqrt{\frac{2}{2+\xi}}\lp 1+\xi \rp^{1/4}$.  For a $\xi=0.4$ polarization difference one finds $F=0.99$.

 At long distances, these approximate interaction potentials are exact for dipoles aligned by a
 (perhaps very small) electric field, projecting the dipolar interaction into the lowest band,
  but at short distances it is just a convenient approximation.   The form is chosen so that at short distances the potential goes to a constant, as does the true projected potential, and approximately interpolates between the exact long-distance and zero distance forms.
  In fact, we will set $A_{\alpha\beta}=0$.  Although for
  some calculations this can lead to divergent results, everything in this manuscript remains finite in the $A_{\alpha\beta}\rightarrow 0$ limit,
  and comparing the $A_{\alpha\beta}\rightarrow 0$ limit with alternative calculations using finite
  realistic values of $A_{\alpha\beta}$  (not presented) shows excellent agreement.
The fundamental reason for this is that at low temperature and for a dilute gas, the physics is  fairly insensitive to the short-range physics governed by $A_{\alpha\beta}$.

 The $p$-wave interaction Hamiltonian is (the multi-internal state generalization of~\cite{idziaszek:analytical_2009}, as given in Ref.~\cite{lemke:p-wave_2011})
\be
H_{\text{pw}} &=& H_{\text{pw}}^{(11)}+H_{\text{pw}}^{(12)}+H_{\text{pw}}^{(21)}
\ee
where
\be
H_{\text{pw}}^{(\alpha\beta)} &=& u_{\alpha\beta} W_{\alpha\beta} \sum_i \int\!d\v{\rho}\bigg[\lp\nabla_\rho \psi^\dagger_{\alpha}(i,\v{\rho})\rp \psi_{\beta}^\dagger(i,\v{\rho})\nonumber \\
    &&\hspace{1.2in}{}-\psi^\dagger_{\alpha}(i,\v{\rho}) \lp \nabla_\rho \psi^\dagger_{\beta }(i,\v{\rho})\rp  \bigg]\nonumber \\
    &&\hspace{-0.3in}{}\cdot \lb \lp \nabla_\rho \psi_{\beta}(i,\v{\rho})\rp \psi_{\alpha }(i,\v{\rho})- \psi_{\beta}(i,\v{\rho}) \lp \nabla_\rho \psi_{\alpha}(i,\v{\rho})\rp \rb
\ee
with $u_{\alpha\beta}= 6\pi \hbar^2 b_{\alpha\beta}^3/m $, $W_{\alpha\beta}=\int \! dz\, |w_{\alpha}(z)|^2|w_{\beta}(z)|^2$, and  $b_{\alpha\beta}$ the $p$-wave scattering length between states $\alpha$ and $\beta$.  In asymptotically deep lattices the Wannier functions are approximately Gaussian, and $W_{\alpha\beta}$ simplifies to $
W_{\alpha \beta} \doteq \frac{\sqrt{\pi}}{d} \lp\sqrt{\frac{1}{{\bar V}_\alpha}}+\sqrt{\frac{1}{{\bar V}_\beta}}\rp^{-1/2}
$
with ${\bar V}_\alpha$ the lattice depth in recoil units for state $\alpha$. If the lattice potentials for $\alpha$ and $\beta$ are the same, then $W_{\alpha\beta} \doteq (1/d) \sqrt{\pi/2} {\bar V}^{1/4}$.
Alternatively, the $p$-wave interaction could be written by adding a term to the intermolecular potential $V_{\alpha \beta}$ in Eq.~\eqref{eq:proj-dipole-1} of the form $(2\pi b_{\alpha\beta}^3/m) {\buildrel\leftarrow\over \nabla_\rho} \delta(\v{\rho})\frac{\partial^3}{\partial \rho^3}\rho^3  {\buildrel\rightarrow \over \nabla_\rho}\int \!dz\, |w_{\alpha}(z)|^2|w_{\beta}(z)|^2$~\cite{idziaszek:analytical_2009}, where an arrow to the left indicates that the gradient operates to functions on the left.

There is also a confining potential
\be
H_{\text{trap}} &=& \sum_{i,\alpha} \int\!d\v{\rho}\, V_\alpha(i,\v{\rho}) \psi^\dagger_{\alpha}(i,\v{\rho})\psi_{\alpha}(i,\v{\rho}),
\ee
which we neglect in our first computation of the average spectral shifts for the homogeneous system in Secs.~\ref{sec:general-sumrule} and~\ref{sec:weakly-interacting-sumrule}.  Its effects will be included with a local density approximation in Sec.~\ref{sec:sumrule-lda}.

One measures the recoil-free spectrum by applying a probe pulse for a short time, where the probe pulse is described by the Hamiltonian
\be
H_p &=& \Omega\sum_{\v{j}} \int\!d\v{\rho}\, e^{-i \omega  t}\psi_{1}(j,\v{\rho})\psi_{2}^\dagger(j,\v{\rho})+\hc
\ee
in the rotating wave approximation, with $\omega$ the probe frequency and $\Omega$ proportional to the amplitude of the probe field.
The spectrum is given by counting the number of atoms transferred to state $2$ as a function of $\omega$ for a fixed probe time.

As an aside, we note that in atomic clock experiments, the small variations of $\Omega$ with the trap quantum number give rise to a shift even for $s$-wave interacting -- initially identical -- fermions~\cite{rey:many-body_2009,yu:clock_2010}.  
This effect is quite tiny, coming from the differences in Rabi frequencies associated with each trap mode: they only manifest in clocks because of the clock experiments' extraordinary frequency resolution ($\sim 1$Hz).
The effect is even smaller  for current experiments: the variation of this Rabi frequency for radio-frequency or microwave pulses is orders of magnitude smaller, and so the resulting shift will be much smaller than a Hertz.  This is completely negligible in the current experiments.

Finally, we comment that we calculate all frequency shifts relative to the level splitting of the internal states (e.g. rotational levels) 1 and 2 of a single molecule, $\delta_v$.  The experimentally measured spectra will then all be shifted by this value.

\section{Spectral shift: sum-rules}

\subsection{General expressions for the homogeneous gas \label{sec:general-sumrule}}

The spectral weight for the recoil-free spectrum is the number of particles transferred from the initial internal state in a time $t$, as defined above. Fermi's Golden Rule  gives
\be
{\mc I}(\omega) &=&\frac{2\pi t}{\hbar} \sum_{i,f} p_i |\opsandwich{f}{H_p}{i}|^2\delta(\omega-E_f+E_i),\label{eq:FGR-RF-spectra}
\ee
where the sum over $i$ and $f$ runs over all possible energy eigenstates of $H$, $p_i$ is the initial probability of being in state $i$, and $E_j$ is the energy of state $j$ for the Hamiltonian $H$.

The average spectral frequency
\be
\expec{\omega} &=& \frac{\int \! d\omega\, \omega {\mc I}(\omega)}{\int \! d\omega\, {\mc I}(\omega)}
\ee
is~\cite{oktel:cs-ref2,oktel_optical_1999}
\be
\expec{ \omega} &=& \frac{\expec{[H_p,H]H_p}}{\expec{H_p^2}},\label{eq:comm-sum-rule}
\ee
where the expectation includes both the thermal and quantum averages.
This is an \textit{exact} result (within linear response) relating $\expec{\omega}$ to the expectation value of an operator which can simply be evaluated from the definitions of $H_p$ and $H$, for \textit{any} Hamiltonian, and holds even for strongly correlated systems.

We evaluate the commutators in Eq.~\eqref{eq:comm-sum-rule} for each interaction term in $H$ in turn.  The spectral shift from $s$-wave interactions vanishes.

The spectral shift from the direct and spin flip terms of the dipolar potential is
\be
\expec{\omega_d}\! &=&\! n \sum_j \int \! d\v{\rho} \lb V_{12}(j,\v{\rho}) - V_{11}(j,\v{\rho})+V_{\text{sf}}(j,\v{\rho})\rb \nonumber \\
    &&\hspace{0.5in}{}\times g_2(j,\v{\rho}), \label{eq:gen-shift-dipole}
\ee
with
\be
n&=&\expec{\psi^\dagger_1(0,\v{0})\psi_1(0,\v{0})}\label{eq:2d-dens-1}
 \ee
 the two-dimensional density and
\be
g_2(i,\v{\rho})\! &=&\! \frac{1}{n^2} \expec{\psi^\dagger_1(i,\v{\rho})\psi^\dagger_1(0,\v{0})\psi_1(0,\v{0}) \psi_1(i,\v{\rho}) }\!.
\ee
This holds for any two-body potential, and one sees that it vanishes for an $s$-wave contact interaction because $\expec{\psi^\dagger(0)\psi^\dagger(0)\psi(0)\psi(0)}=0$.  The key qualitative features of this result are understood by mean field theory: the expression,  up to constant factors in each term, is the mean field interaction energy of the state with one molecular state changed (a superposition of states with a single molecular state changed at position $\v{\rho}$, summed over $\v{\rho}$, giving $V_{12}$ and $V_{\text{sf}}$ terms) minus the state with all the molecules in the same initial state ($V_{11}$ term).


The spectral shift for the $p$-wave potential is
\be
\expec{\omega_{pw}} &=& \frac{8\lp u_{12}W_{12}-u_{11}W_{11}\rp}{n}\nonumber \\
    &&\hspace{-0.35in}{}\times
    \expec{\lp \nabla \psi_1^\dagger(0,\v{\rho}) \rp n_1(0,\v{\rho}) \cdot\lp \nabla\psi_1(0,\v{\rho})\rp} \label{eq:gen-shift-pwave}
\ee
with $n_1(j,\v{\rho}) = \psi^\dagger_1(j,\v{\rho})\psi_1(j,\v{\rho}) $.
The total average spectral shift is $\expec{\omega} = \expec{\omega_d}+\expec{\omega_{pw}}$.

\subsection{Weakly interacting homogeneous gas \label{sec:weakly-interacting-sumrule}}

Equations~\eqref{eq:gen-shift-dipole} and~\eqref{eq:gen-shift-pwave} give exact, general relations for the spectral shift, valid in any system.  Here we simplify these for weakly interacting systems.  In the limit where the interactions induce negligible correlations, the system's correlations are that of a free Fermi gas.  In other words, since we want to know the shifts to lowest order in the interactions, and the interaction already appears multiplying the operator expectation values, we may evaluate these expectation values to zero'th order in the interaction.  Consequently Wick's theorem determines the expectation values.

The spectral width of a homogeneous gas narrows as the interactions decrease, as we show in Sec.~\ref{sec:collisional-broadening}.  For experiments considered in the present manuscript, the spectral width is very small and the spectrum is nearly a perfect $\delta$-function.
In this case, the sum rule gives the exact location of the well-defined spectral peak.

\subsubsection{Expressions at general temperature}

For the dipolar interaction, we first compute $g_2$.  Wick's theorem gives
\be
g_2(i,\v{\rho}) &=& 1-\lp\frac{G(i,\v{\rho})}{n}\rp^2\label{eq:g2-general}
\ee
where $G(i,\v{\rho})=\expec{\psi^\dagger_1(i,\v{\rho})\psi_1(0,\v{0})}$.
Because there is no tunneling, the Green's function has the form
\be
G(i,\v{\rho}) &=& \delta_{i,0} {\bar G}(\v{\rho})
\ee
with
\be
{\bar G}(\v{\rho}) &=& \frac{1}{(2\pi)^2} \int\!d\v{k}\, e^{i\v{k}\cdot \v{\rho}}\expec{{\hat n}_{1k}},\label{eq:barg-1}
\ee
where the integral runs over all the transverse momentums states of the system (which, recall, in this section is assumed to be homogeneous).

The expectation values are, from Eq.~\eqref{eq:2d-dens-1},
\be
n &=& \frac{1}{2\pi} \int \! k dk \, \frac{1}{e^{\beta(k^2/(2m)-\mu)}+1}  \nonumber \\
    &=& \frac{m}{2\pi \beta} \log \lp 1+e^{\beta \mu}\rp
\ee
and, from Eq.~\eqref{eq:barg-1}
\be
{\bar G}(\rho) &=& \frac{1}{2\pi} \int \! k dk \, \frac{J_0(k\rho)}{e^{\beta(k^2/(2m)-\mu)}+1} \label{eq:barG-general}
\ee
with $J_0$ the zero'th Bessel function of the first kind.
Equations~\eqref{eq:gen-shift-dipole},~\eqref{eq:g2-general}, and~\eqref{eq:barG-general} give the dipolar shift.

For the $p$-wave interaction, Wick's theorem gives
\be
\expec{\omega_{pw}} &=& \frac{8}{n} \lp u_{12}W_{12}-u_{11}W_{11}\rp
\nonumber \\
    &&\hspace{-0.3in}{}\times
    \expec{\lp\nabla \psi_1^\dagger(0,\v{\rho})\rp\cdot \lp \nabla \psi_1^{\phantom{\dagger}}(0,\v{\rho}) \rp }\expec{n(0,\v{\rho})} \nonumber \\
    &=& \frac{8}{(2\pi) } \lp u_{12}W_{12}-u_{11}W_{11}\rp\nonumber \\
    &&\hspace{0.1in}{}\times \int_0^\infty\! dk  \frac{k^3}{e^{\beta \lb k^2/(2m)-\mu\rb}+1} \nonumber \\
    &=& -\frac{8m^2 T^2 }{\pi } \lp u_{12}W_{12}-u_{11}W_{11}\rp \text{Li}_2(-e^{\beta \mu})\nonumber \\
    \label{eq:shift-pwave-gen}
\ee
where $\text{Li}_\alpha$ is the polylog function.

\subsubsection{Thermal (Boltzmann) and degenerate gas limits}

In this section we evaluate the shifts, given above for a weakly interacting system at arbitrary temperature, in the thermal/Boltzmann ($T\gg \mu$) and degenerate ($T\ll \mu$) limits.

In the thermal limit $T\gg\mu$ the dipolar shift, given by Eqs.~\eqref{eq:gen-shift-dipole},~\eqref{eq:g2-general}, and~\eqref{eq:barG-general} taking $A_{\alpha\beta}=0$,  is
\be
\expec{ \omega_{d}} &=& n \lp \gamma_{12}-\gamma_{11}+\eta\rp  \bigg[ \int \!d\v{\rho}\, \frac{1-e^{-m\rho^2/\beta}}{\rho^3} \nonumber \\
    &&\hspace{0.35in}{}+  2\sum_{j=1}^\infty \int\!d\v{\rho}\, \frac{1-3\lp \frac{(jd_l)^2}{\rho^2+(jd_l)^2} \rp }{(\rho^2+(jd_l)^2)^{3/2}}\bigg]\nonumber \\
    &=& 2\pi^{3/2} \sqrt{\frac{m}{\beta}} n \lp \gamma_{12}-\gamma_{11}+\eta\rp
    \hspace{0.15in} \text{thermal},\label{eq:dipole-shift-therm}
\ee
where $n = [m/(2\pi \beta)] e^{\beta \mu}$.
Note the integrals in the sum over $j$ vanish.  The physical reason for this is that these integrate the electric field generated by the zero'th pancake over the plane of  pancake $j$.  Closing this planar surface at infinity, the total electric flux penetrating the surface must be zero. Since the dipolar field decays as $1/r^3$ the contribution from the surface at infinity is negligible, the electric flux through the plane must be zero, and hence the integral vanishes.

In the degenerate limit $T\ll \mu$, the dipolar shift is
\be
\expec{\omega_{d}} &=& n \lp \gamma_{12}-\gamma_{11}+\eta\rp  \lb \int \!d\v{\rho}\, \frac{1-4\frac{\lb J_1(k_F \rho)\rb^2}{(k_F \rho)^2}}{\rho^3} \rb\nonumber \\
&=& \frac{256}{45} nk_F  \lp \gamma_{12}-\gamma_{11}+\eta\rp \hspace{0.15in}\text{degenerate} \label{eq:dipole-shift-deg}
\ee
with $k_F\equiv\sqrt{2m\mu}$ the Fermi wavevector, and $n= k_F^2/(4\pi)$ in this limit.

The $p$-wave shift, Eq.~\eqref{eq:shift-pwave-gen} simplifies in the thermal $T\gg \mu$ gas to
\be
\expec{\omega_{pw}} &=&  \frac{8 mT}{\pi}n\lp u_{12}W_{12}-u_{11}W_{11}\rp \hspace{0.1in}\text{thermal}\label{eq:thermal-pwave-shift}
\ee
and in the degenerate $T\ll \mu$ gas to
\be
\expec{\omega_{pw}} &=&  \frac{k_F^4}{\pi} \lp u_{12}W_{12}-u_{11}W_{11}\rp \hspace{0.1in}\text{degenerate.}
\ee

\subsection{Spectra in a trap \label{sec:sumrule-lda}}

In this section we calculate the spectra of systems in a harmonic trap.  We use the local density approximation (LDA): we approximate the properties at a position $(i,\v{\rho})$ in the trap by those of a homogeneous system at an effective chemical potential $\mu_{\text{eff}}(j,\v{\rho})=\mu_0-(m\omega_t^2/2)\lb (jd_l)^2+\rho^2\rb$ with $\mu_0$ the global chemical potential of the system and $\omega_t$ the harmonic oscillator potential's trapping frequency.  As usual, the LDA is accurate when the local chemical potential variation over the correlation lengths is sufficiently small~\cite{pethick:p-s-book}. This condition is met in the experiments of present interest.  There is an additional requirement for the accuracy of the LDA, because we are considering a rather nonlocal quantity, the shifts from the the dipolar interaction energy. For this quantity, accuracy of the LDA requires that the spatial variation of the local chemical potential is sufficiently slow compared to the length over which the dipolar interaction becomes negligible.  Here, the accuracy in a spatially varying system will \textit{not} be expected to converge exponentially, but rather the error $\epsilon$ scales as a power law $\epsilon\sim n\int_L^\infty \!(2\pi\rho)d\rho\frac{1}{\rho^3} \propto  n/L$ in the scale $L$ over which $\mu$ varies ($L$ is on the order of the cloud width).  Meanwhile, as seen in Eq.~\eqref{eq:gen-shift-dipole}, the contributions to the shift scale as $n/\ell$ with $\ell$ a characteristic microscopic scale (thermal de Broglie wavelength in the thermal gas or Fermi wavelength in the degenerate gas).  Since (for our trap parameters) $L$ is typically much larger than tens of micrometers, while  $\ell$ is typically a tenth of a micrometer, we expect this to be quite accurate.  However, it will lead to quantitative corrections at the $\lsim 1\%$ level.
Consequently the LDA is a good approximation for systems with many particles, and is typically an excellent approximation for cold atomic systems. Here we take the trap to be harmonic and isotropic for simplicity, but these restrictions are straightforwardly relaxed.

 Within the LDA, the trap summed spectrum is
\be
{\mc I}(\omega) &=& 2\pi \Omega^2 t\sum_i \int \! d\v{\rho}\, n[\mu_{\text{eff}}(i,\v{\rho})]\delta\lb \omega-{\omega}_{\mu=\mu_{\text{eff}}(i,\v{\rho})} \rb\!.\nonumber \\
\label{eq:LDA-gen-spectra}
\ee
We can generally numerically sample the integral, approximating the delta function as a sharply peaked Lorentzian of width $\Gamma$, $\delta(x) \rightarrow \delta_\Gamma(x) $ with $\delta_\Gamma(x) \equiv (1/\pi)\Gamma/(x^2+\Gamma^2)$.  In some cases, however, an analytic formula can be derived, as we now discuss.

The spectral weight in the frequency window $(\omega,\omega+d\omega)$, given by $\mc I(\omega)d\omega$, is $2\pi\Omega^2 t$ times the number of atoms with spectral shifts in that window:
\be
{\mc I}(\omega) &=&  (2\pi\Omega^2 t) 4\pi [r(\omega)]^2 n(r(\omega)) |dr/d\omega| \label{eq:LDA-general-spectrum-dens-propto-number}
\ee
where $r(\omega)$ is the distance to the trap center of atoms that have  spectral shift equal to $\omega$.   Within the LDA, $\expec{\omega}$ as a function of $r$ is inverted to give $r(\omega)$ and one similarly can calculate $d\omega/dr$.

In the thermal gas we obtain analytic formulas.  Note that in this regime the sole chemical potential dependence of both the $p$-wave and dipolar interactions comes from the density, which multiplies a term independent of the chemical potential: $\expec{\omega}= C n$ for some constant $C$.  Moreover, in this limit, $n(r) = [m/(2\pi \beta)] e^{\beta \mu_{\text{eff}}(r)}$. Then $r(\omega) = (1/\omega_t)\sqrt{2/(m\beta)}\sqrt{\beta \mu_0 - \log\lp 2\pi \beta \omega/(Cm)\rp} $ and $|dr/d\omega|=\lb\sqrt{2m\beta}\omega\omega_t \sqrt{\beta \mu_0 - \log \lp 2\pi \beta \omega/(Cm)\rp}\rb^{-1}$.
Thus the trap summed spectrum of the thermal gas is
\be
{\mc I}(\omega) &\propto&
\sqrt{\beta \mu_0-\log \lp \frac{2\pi \beta \omega}{C m}\rp}\hspace{0.2in} \text{trap, thermal}\label{eq:trapped-thermal-spectra-analytic}
\ee
for $\omega$ such that the term under the square root is real and positive, and zero otherwise.

No simple expression exists for the case of general temperature.  Nor does one exist even if the gas is deeply degenerate, although we mention that such formulas \textit{do} exist for the deeply degenerate gas if \textit{either} the $p$-wave or dipolar shift is included alone.  For a purely dipolar gas in the deeply degenerate regime ($T=0$), one finds
\be
{\mc I}(\omega) &\propto&
\omega^{1/3} \sqrt{\frac{\mu_0}{\omega_{\text{deg}}}-
\lp\frac{\sqrt{2}\pi\omega}{\omega_{\text{deg}}}\rp^{2/3}} \hspace{0.05in} \text{trap, degenerate}\nonumber \\
&&\label{eq:trapped-deg-spectra-analytic}
\ee
with $\omega_{\text{deg}}\equiv \lb m^3 \lp \frac{256}{45}\lp \gamma_{12}-\gamma_{11}+\eta\rp\rp^2\rb^{-1}$.

\subsection{Example spectra\label{sec:example-spec}}

\begin{figure}[hbtp]
\setlength{\unitlength}{1.0in}
\includegraphics[width=3.1in,angle=0]{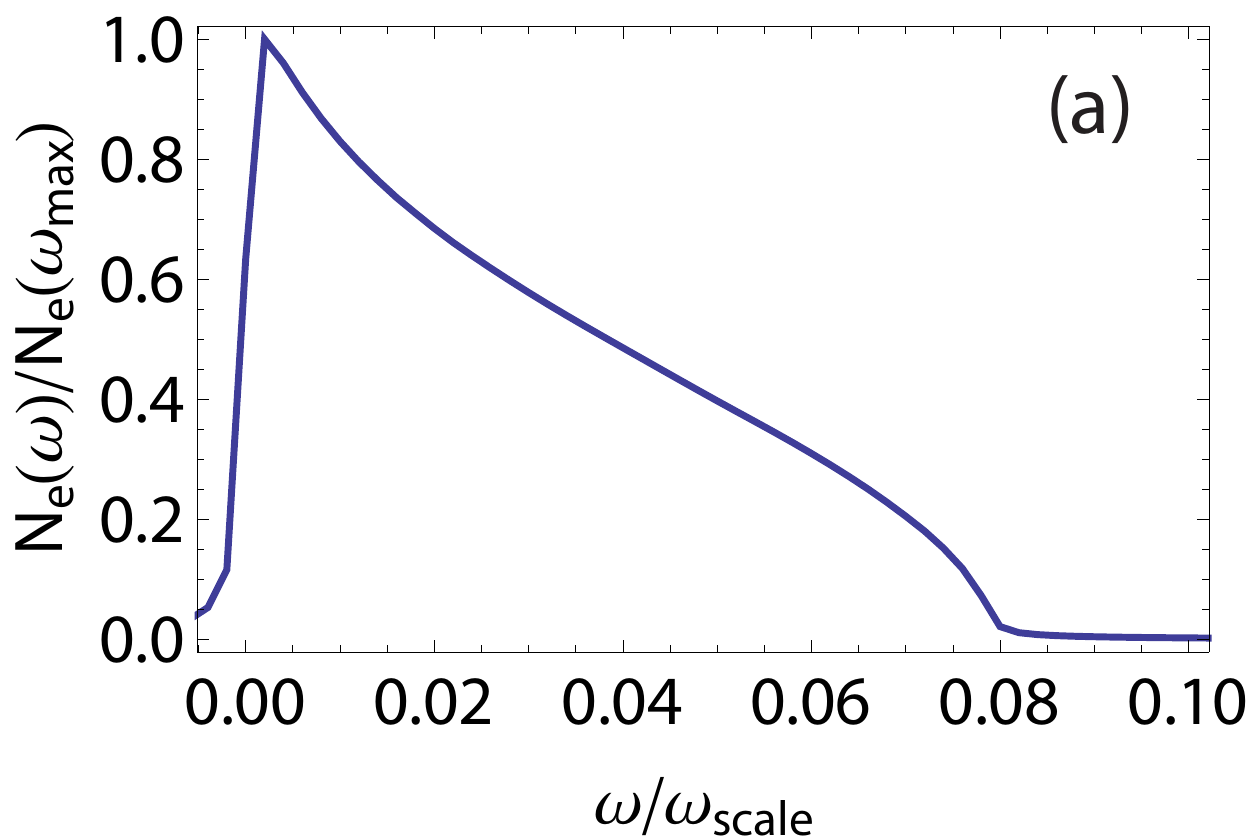}
\includegraphics[width=3.08in,angle=0]{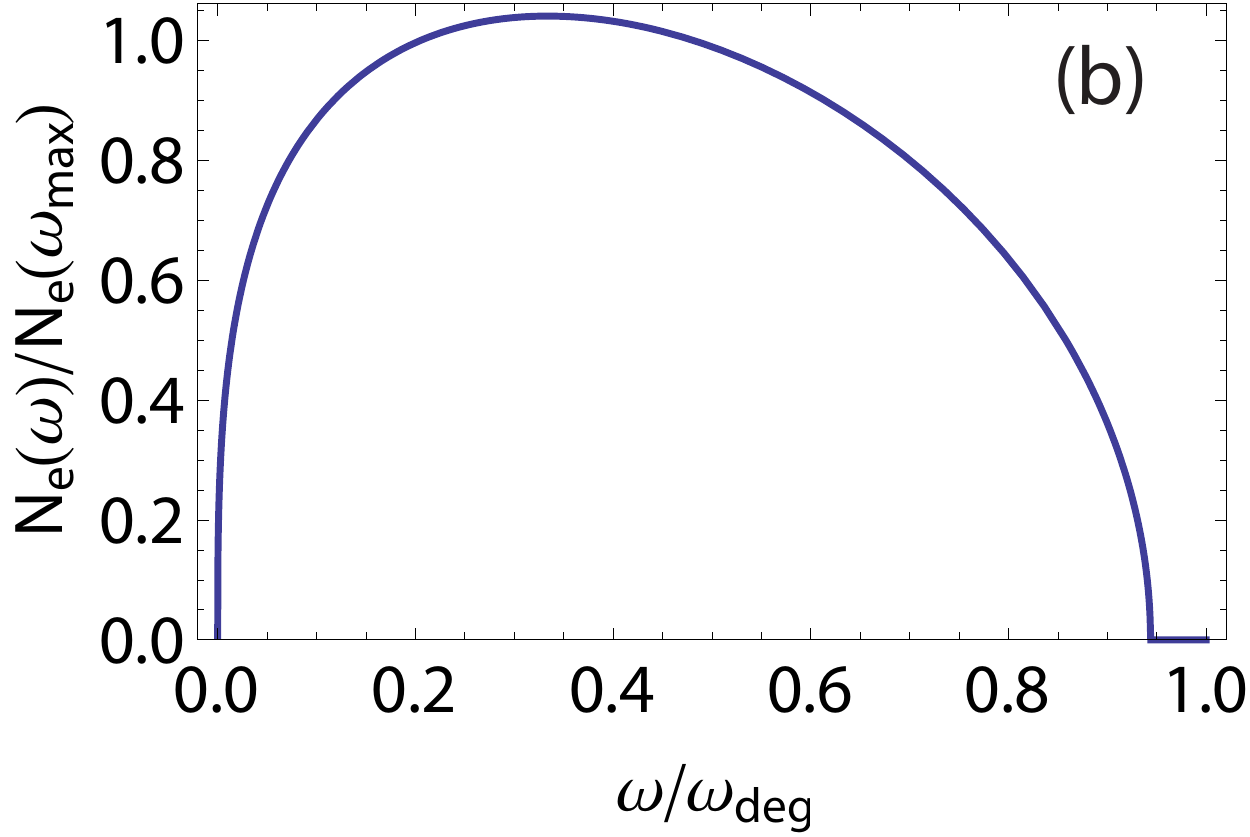}
\caption{(a) Recoil-free spectrum with a rough modeling of the JILA experimental parameters. Here $N_e(\omega)/N_e(\omega_{\text{max}})$ is the number of excited particles after a probe pulse of frequency $\omega$ and duration $t$ rescaled by the peak's maximum value. The overall frequency scale, related to the dipolar interaction strengths, is $\omega_{\text{scale}}=\lp\gamma_{12}-\gamma_{11}+\eta\rp\sqrt{\frac{\pi m^3}{\beta^3}}$.  The \textit{shape} of this spectrum is entirely determined by the density. The infinite system has a logarithmically divergent peak at $\omega=0$, but this is rounded off by a small spectral broadening $\Gamma=7\times 10^{-4}\omega_{\text{scale}}$ added to the calculation here and, less so, by the finiteness of the present system, assumed to be 200 lattice sites wide.  (b) Recoil-free spectrum at $T=0$ (deeply degenerate) with $\omega_{\text{deg}}\equiv \lb m^3 \lp \frac{1024}{45}\lp \gamma_{12}-\gamma_{11}+\eta\rp\rp^2\rb^{-1}$, where we have chosen $\mu_0= 2.6 \omega_{\text{deg}}$; different $\mu_0$'s just rescale the plot.
 \label{fig:trapped-spec-2d-nondeg}}
\end{figure}

Figure~\ref{fig:trapped-spec-2d-nondeg}~(a) shows an example spectrum for a thermal gas calculated by numerically integrating the LDA expression Eq.~\eqref{eq:LDA-gen-spectra} with the dipolar shifts Eq.~\ref{eq:dipole-shift-therm}, for a finite size system of 200 pancakes (although only $\sim30$ of these are significantly occupied) and trapping frequencies and temperatures roughly corresponding to ongoing JILA experiments: $\omega_t=2\pi \times 100$~Hz  and $T=800$~nK, with 2200 particles in the central tube.  In fact, the JILA traps are quite anisotropic, but, since our goal here is not to construct detailed models of particular experiments, an isotropic trap gives an adequate caricature.  Fig.~\ref{fig:trapped-spec-2d-nondeg}~(b) shows the spectrum at $T=0$, showing how the spectral shape changes when the gas becomes deeply degenerate.   In contrast to the asymmetric thermal cloud with its small-$\omega$ logarithmic divergence, a more symmetric, broad peak forms. At intermediate temperatures, the spectrum looks roughly like a sum of these two spectra: there is a broad spectral peak from the degenerate portion of the trap and a small-$\omega$ logarithmic divergence.
This shows how the spectra may be used to assess the degeneracy of the gas.

The structure shown in Fig.~\ref{fig:trapped-spec-2d-nondeg} is quite general.  
A trapped system's spectra generically shows two characteristic features: a logarithmic divergence at low frequency due to thermal tails and a broad structure when the gas becomes strongly interacting and/or degenerate~\cite{mueller:generic_2008}.
Such structure has also been seen in other systems, such as dilute spin-polarized atomic hydrogen~\cite{fried:hydrogen}.

\begin{figure}[hbtp]
\setlength{\unitlength}{1.0in}
\includegraphics[width=3.3in,angle=0]{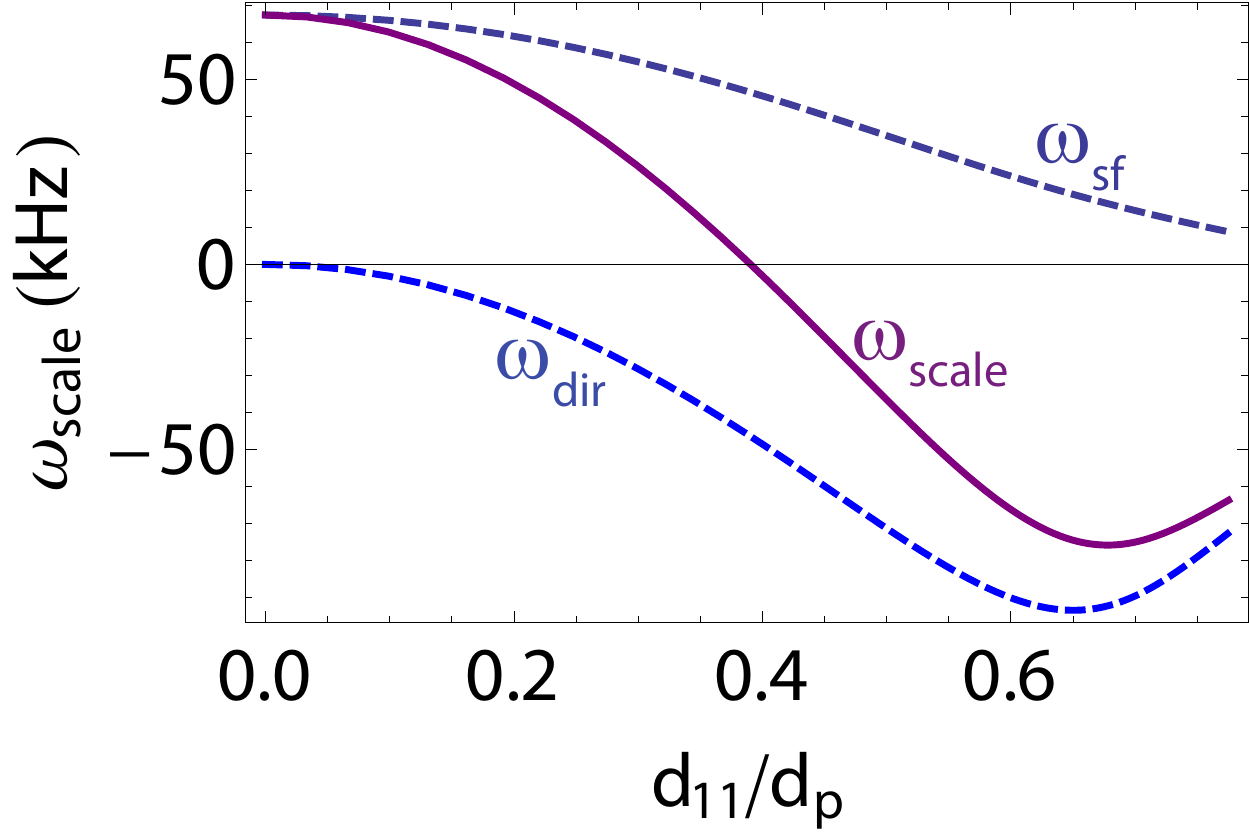}
\caption{The characteristic frequency of the collisional shifts in the thermal gas, $\omega_{\text{scale}}$, for transitions between KRb's two lowest rotational states versus the dipole moment $d_{11}$ in units of the permanent dipole moment $d_p$ (roughly 0.5 Debye for KRb). Middle curve: total characteristic frequency $\omega_{\text{scale}}=\omega_{\text{sf}}+\omega_{\text{dir}}$.  Top curve: $\omega_{\text{sf}}$, the contribution from spin-flip term. Bottom curve: $\omega_{\text{dir}}$, the contribution from the direct term.
\label{fig:omega-scale}}
\end{figure}

The dipolar energy scale  $\omega_{\text{scale}}=\omega_{\text{dir}}+\omega_{\text{sf}}$, with $\omega_{\text{dir}}\equiv(\gamma_{12}-\gamma_{11})\sqrt{\frac{\pi m^3}{\beta^3}}$ and $\omega_{\text{sf}}\equiv \eta\sqrt{\frac{\pi m^3}{\beta^3}}$
depends on the rotational states being used and the external electric field applied.  For concreteness, we estimate the shift in the transition from the ground rotational state $\ket{0,0}$ to the excited rotational state  $\ket{1,0}$.  The magnitudes of shifts for other low-lying rotational transitions are generically  the same order of magnitude.  Fig.~\ref{fig:omega-scale} plots the energy scale  $\omega_{\text{scale}}$ versus the dipole moment in units of the permanent dipole moment.  The permanent dipole moment for KRb is about 0.5 Debye.  We calculate the dipolar interaction and $\omega_{\text{scale}}$ by computing the dipole moments of these rotational states using a quantum rigid rotor model in an electric field with parameters given in Ref.~\cite{kotochigova:electric_2010}.  We observe that this energy scale can be greater than 50 kHz.
Then we see from Fig.~\ref{fig:trapped-spec-2d-nondeg}~(a) that even for the low densities of the JILA experiments the width of the spectra is as large as about 5kHz.  Experimentally resolving such scales is standard practice.

The real part of the $p$-wave interaction is unknown, but it should be quite small  at these low temperatures.  On the other hand, the imaginary part of the $p$-wave scattering, due to the reactive loss when molecules collide, can be much faster, and limits the cloud lifetime for KRb, even when it is suppressed in by using a  two-dimensional ``pancake" geometry~\cite{miranda:controlling_2011}.  It gives an imaginary shift to the average frequency from the sum rule, and leads to a Lorentzian broadening.
To determine the value of this broadening, note that Eq.~\eqref{eq:thermal-pwave-shift} with $b_{12}=0$ is the expectation value of the $p$-wave interaction energy.  The imaginary part is the particle lifetime, since at the level of calculation of this equation the imaginary part of this energy is the quasiparticle lifetime and the only source of quasiparticle decay is the $p$-wave particle loss.     Thus the cloud lifetime is simply this quasiparticle lifetime.
Using cloud lifetimes measured in Ref.~\cite{miranda:controlling_2011} that decay is around $4/s$, or  a little less than a Hertz.  Since the shift is the difference of two such rates -- the $1$-$2$ channel and the $1$-$1$ channel -- we expect it to be at most on this order, and likely considerably less.  Of course we can't rule out the possibility that the $1$-$2$ interaction is much larger, but it is highly unlikely to be the orders of magnitude larger required to contribute substantially to the spectral broadening, compared to the dipolar shifts.  Thus the $p$-wave broadening   will be negligible under the current conditions of the JILA experiments.

\section{Spectral shape: broadening}

Previously, we considered the mean spectral shifts. Also, because the $p$-wave shifts may have an imaginary part due to the reactive nature of certain molecules, such as KRb, this gives a broadening.
 Here we go beyond treating the spectra as sharp delta functions with this simple broadening, and  consider the spectral lineshape for the experimentally relevant dilute gas.  Broadening arises from several sources:  differing trapping potentials felt by the two states due to their differing polarizabilities (coming from vector and tensor light shifts),  collisional broadening (related to the quasiparticle lifetime), temperature broadening from going beyond the pseudopotential approximation, and higher band occupation.    We will see that  in this limit all sources of broadening are very small, except the differing state polarizabilities.
 However, even for current conditions, these broaden the spectra by an amount that is comparable to the width of the trap-summed spectra. Consequently some key features survive, such as the maximal shift.  Moreover, by applying the probe after aligning the dipoles at the ``magic angle" or ``magic field strength"~\cite{kotochigova:electric_2010}, even this broadening can be eliminated.

There is also the natural linewidth of the excited rotational state, given by $\Gamma_{\text{nat}} = (\delta_\nu/c)^3 d_{12}^2/(3\pi \epsilon_0 \hbar)$ where $\delta_\nu$ is the transition frequency between the excited state $\ket{1,0}$ and the ground state $\ket{0,0}$, and $d_{12}$ is the dipole matrix element between these states. Even for the strongest dipole moments, this is $\sim 10^{-10}$~Hz for KRb, calculating $d_{12}$ and $\delta_\nu$ in the same manner we calculated the energy scales above, and thus completely negligible.

\subsection{Differing polarizabilities}

The polarizabilities of rotational states can differ substantially for different states due to vector and tensor light shifts.  For example, the polarizability difference of KRb between the lowest and first excited rotational states is about 30\% for lattice wavelengths giving a lattice spacing of 545nm~\cite{kotochigova:electric_2010}.  In systems using optical potentials to confine the molecules, this leads to the molecules experiencing different external potentials.  In the presently considered case, this means the trapping frequencies differ by about 15\%  and the lattice depths differ by about 30\%. Since we are taking the lattice to be sufficiently deep to suppress tunneling, its only effect on physics along the lattice direction is to change the Wannier functions for each state and thus renormalize the interactions. This was accounted for in our derivation of the lattice Hamiltonian. However the differing trapping frequencies in the transverse direction leads to broadening.

Within the LDA, we note that if the trapping potentials differ by $\Delta(\v{r})$ at location $\v{r}$ then this shifts the spectral line of atoms at that position by $s(\v{r})=\Delta(\v{r})$ (this intuitive result follows formally from calculating the sum rule commutators of Eq.~\eqref{eq:comm-sum-rule}). For harmonic traps, this shift is $s(\v{r}) = -A_2+A_1+\frac{m(\omega_2^2-\omega_1^2)}{2}r^2$, where $\omega_\alpha$ is the harmonic oscillator trapping frequency in the transverse direction for state $\alpha$ and $A_\alpha$ are constants. Note that $A_\alpha$ are positive since the optical intensity is maximal at the center of the trap and  that there is a frequency shift there as well.  The magnitude of the shift $A_1-A_2$ depends on the exact trapping laser used, and just gives a constant shift to the spectrum.  What is more important for our present concerns is the broadening coming from the second, spatially-dependent term.  At finite temperatures, there is (exponentially small) occupation arbitrarily far from the trap center, and the shift from the second term increases indefinitely as one moves outwards in the trap.
However, because the spectral density for a given value of this shift is proportional to density, the result of the polarizability difference is a exponentially decaying tail, in frequency, in the spectrum.  The spectrum for locations at distance $r$ from the center of the trap has a polarizability difference shift
\be \omega=\frac{m(\omega_2^2-\omega_1^2)}{2}r^2,\label{eq:pol-diff-freq-position-rel}
\ee
 and the spectral weight is proportional to $n(r)$.  In the thermal gas,
Eq.~\eqref{eq:LDA-general-spectrum-dens-propto-number} with the frequency position relation Eq.~\eqref{eq:pol-diff-freq-position-rel} thus gives the spectrum, at least in the tails where the polarizability difference shift is the only relevant one and the cloud is thermal.  Thus  ${\mc I}(\omega)\propto r^2 (dr/d\omega) e^{-\frac{\beta m\omega_1^2}{2}r^2}$ yielding
\be
{\mc I}(\omega) &\propto&  \sqrt{|\omega|} e^{-\beta \frac{\omega_1^2}{\omega_2^2-\omega_1^2}\omega} \Theta[(\omega_2^2-\omega_1^2)\omega]
\label{eq:pol-diff-spectrum}
\ee
where $\Theta$ is the Heaviside step function.
  The spectral density decays exponentially with frequency,  with the characteristic decay frequency  $[(\omega_2^2-\omega_1^2)T/\omega_1^2]$.
Note that this decreases with decreasing temperature, by virtue of the cloud shrinking so that less particles occupy the wings, where the potential difference is largest.  Also in the tail, this is the only source of shifts since densities are low, so this always gives the complete spectral shape of the spectral tail.

For an 800nK gas with polarizabilities of the two states differing by 30\%, this characteristic frequency scale over which the spectral density decays is about 4.9~kHz, comparable to the spectral width of the trap-summed spectra obtained without this source of broadening.
However,  we emphasize that these shifts can be eliminated by performing the recoil-free spectroscopy while aligning the molecules at the magic angle or at the magic field strength, using states with nearly identical polarizabilities, or using other techniques to compensate the polarizability difference. As such, the spectral tails can be a useful diagnostic of how close the trapping potentials of the two species are, and  can help, for example, to experimentally pinpoint the magic angle.

\begin{figure}[hbtp]
\setlength{\unitlength}{1.0in}
\hspace{0.07in}
\includegraphics[width=2.65in,angle=0]{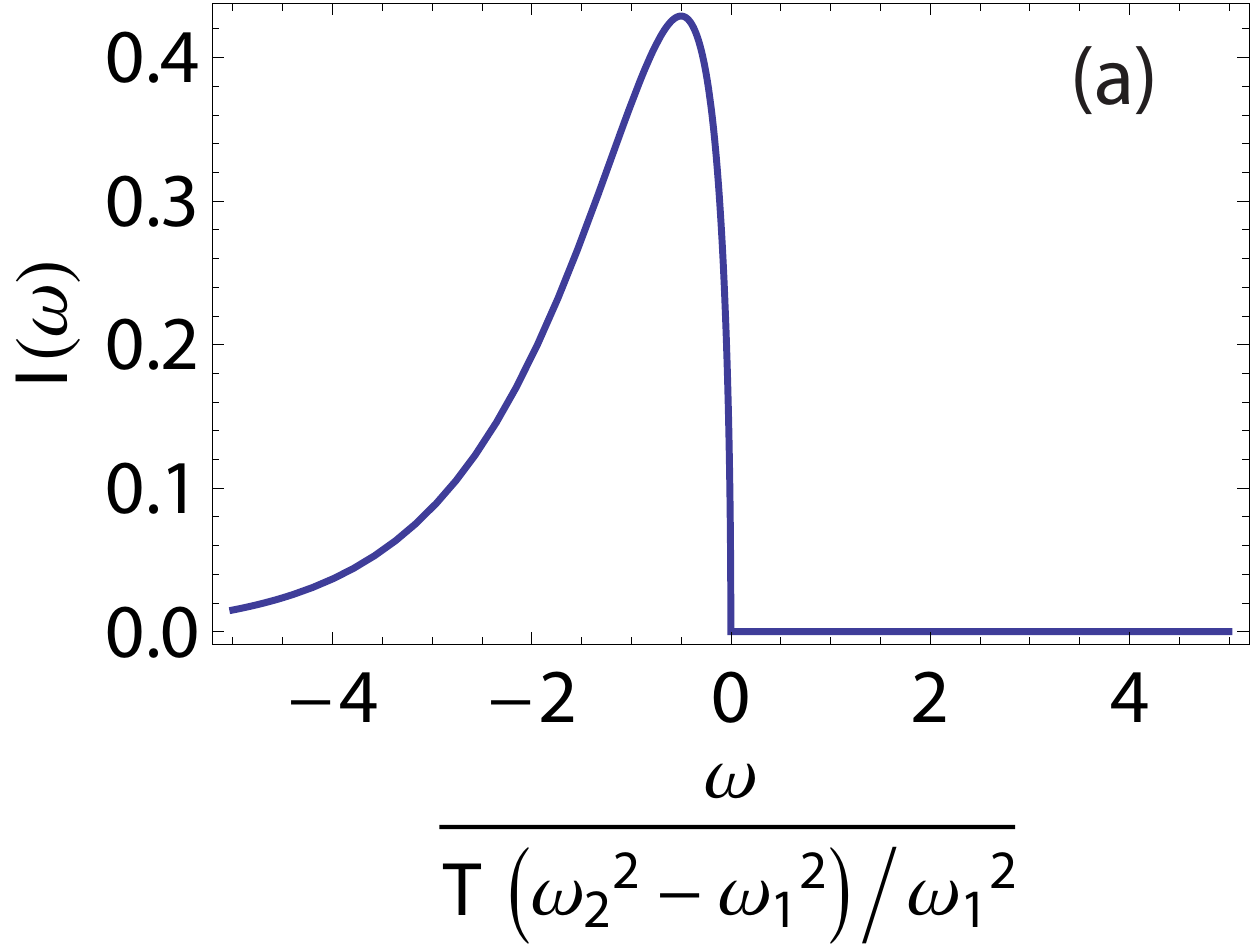}
\vspace{0.07in}
\includegraphics[width=2.9in,angle=0]{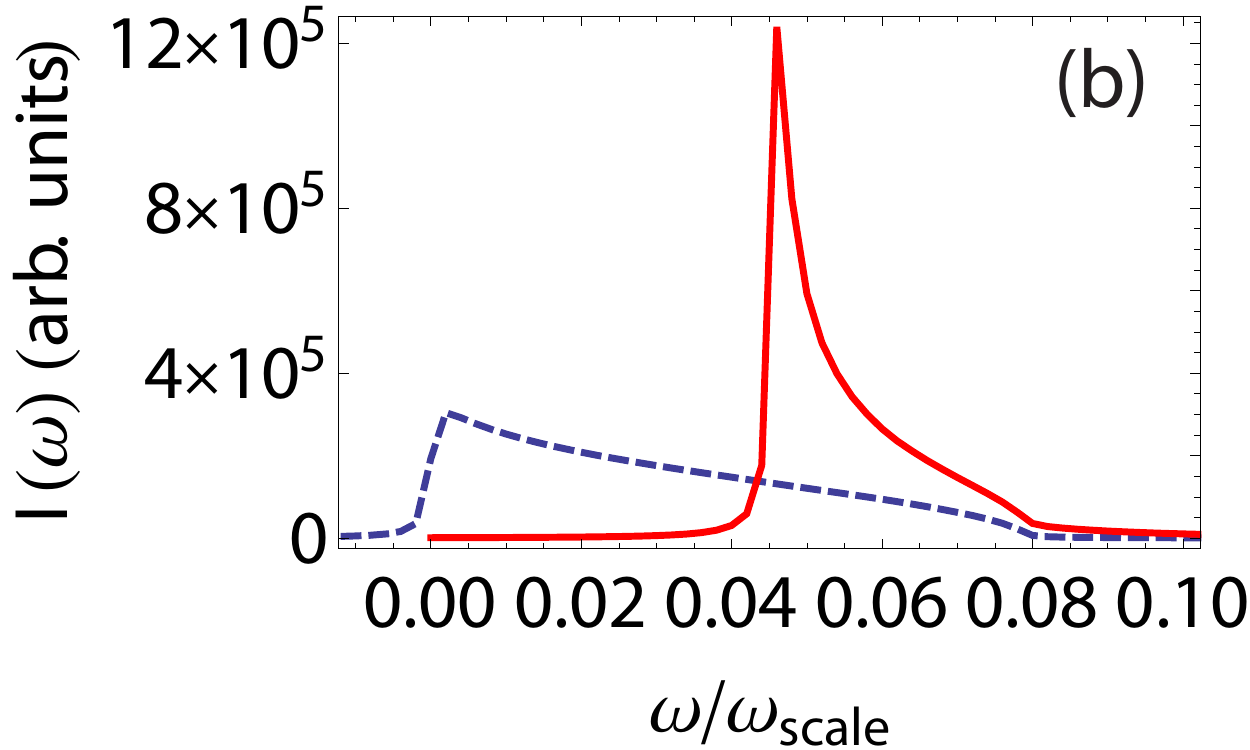}
\vspace{0.05in}
\includegraphics[width=2.7in,angle=0]{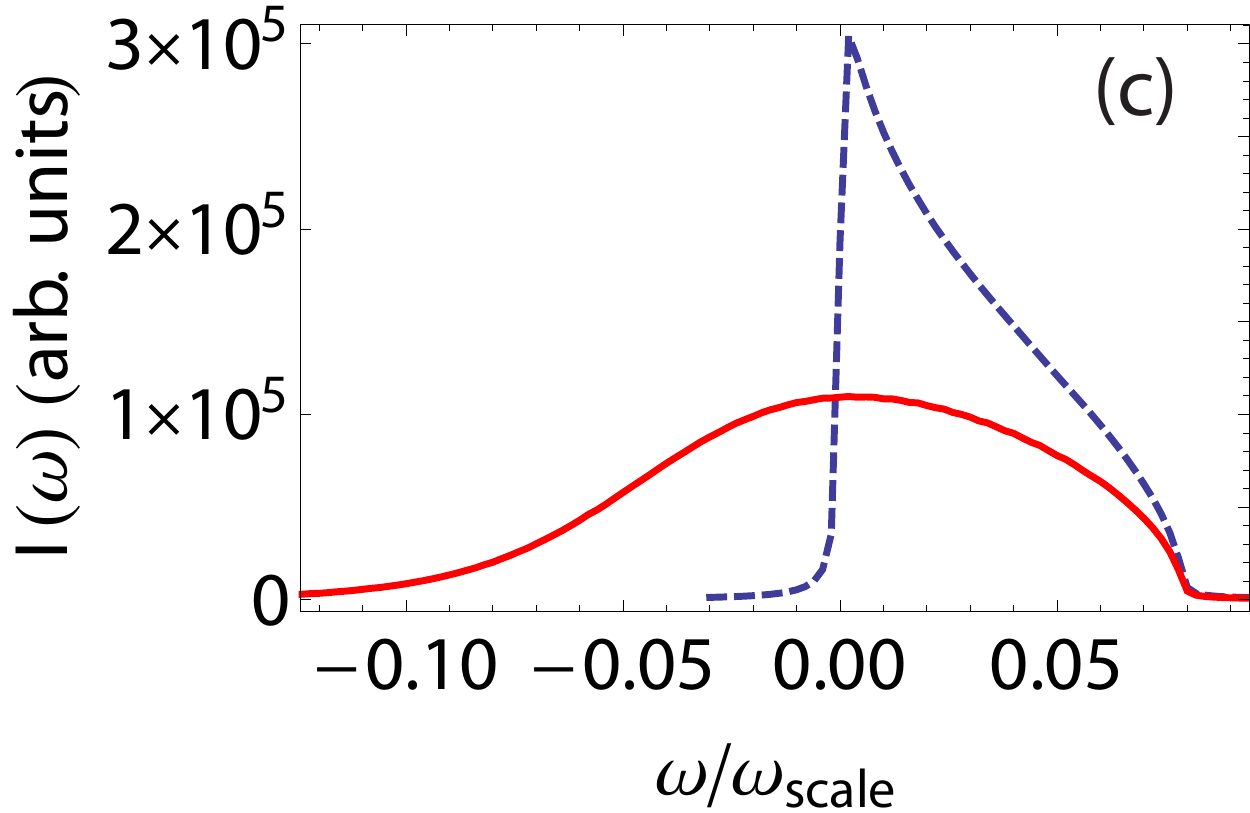}
\caption{
Polarizability difference broadening in the spectra. (a) Polarizability difference broadening when other shifts are negligible (as applies in a very dilute gas, or in the tails of the cloud).
(b,c) Broadening with the polarizability difference  for positive (b) and negative (c) $\omega_2^2-\omega_1^2$, with magnitude  decreased five-fold from its true value.
They compare the thermal spectra without broadening [blue, dashed, exactly as Fig.~\ref{fig:trapped-spec-2d-nondeg}~(a)] with the spectra with broadening [red, solid], for dipole moments giving the maximal values of the collisional shifts relevant for the JILA parameters computed previously.
 \label{fig:pol-diff-broadening}}
\end{figure}

Figure~\ref{fig:pol-diff-broadening}~(a) illustrates the spectral shape due to polarizability difference broadening, with a sufficiently dilute systems such that there is no interaction shift.  In this limit, the spectrum without the polarizability difference broadening would simply be a $\delta$-function.  For illustration, we assumed that $\omega_2^2<\omega_1^2$.  For $\omega_2^2>\omega_1^2$, the spectrum will be reflected across the $y$ axis.  Figs.~\ref{fig:pol-diff-broadening} (b,c) illustrate the effect of this broadening on the spectrum in the presence  of dipolar interaction shifts, analogous to Fig.~\ref{fig:trapped-spec-2d-nondeg}.

\subsection{Collisional lifetime\label{sec:collisional-broadening}}

Interparticle interactions broaden the spectral line by giving the quasiparticle excitations associated with the response a finite lifetime. For weak interactions, the lineshape is roughly Lorentzian.  We emphasize that the full lineshape is captured in Eq.~\eqref{eq:FGR-RF-spectra}.  Only our subsequent determination of trapped system spectra by the LDA and treating the homogeneous systems' spectra as delta functions made an assumption that the linewidth was narrow. Here we will derive the linewidth and show that this assumption is justified under current experimental conditions.
Because we have found the effect of $p$-wave collisions to be substantially smaller than the dipolar collisions, we only consider the latter here.  Moreover, we will treat only the direct term dipolar effects in a single pancake and restrict ourselves to the thermal gas.  Including inter-pancake interactions and spin-flip terms is more tedious, and we will see that the direct, on-site contribution to the broadening -- even at its largest experimentally relevant values  -- is fairly small. Thus, we may expect the other effects -- inter-pancake interactions and spin-flip terms to be small as well.  We return to this in a bit more detail at the end of this section.

We start by calculating the quasiparticle lifetime for particles in the initial rotational state.  The standard theory of collisional lifetimes for a quasiparticle of momentum $\v{p}$ and frequency $\omega$ gives the decay rate~\cite{kadanoff:quantum_1962}
\be
\Gamma(\v{p},\omega) &=& \frac{1}{2(2\pi)^3}\int d\v{p'} d\v{q} \, \delta(\omega+ \epsilon_{\v{p}'} -\epsilon_{\v{p}+\v{q}} - \epsilon_{\v{p'} - \v{q}}) \nonumber \\
&&\hspace{-0.5in}{}\times\lb v(-\v{q}) - v(\v{p}-\v{p}' + \v{q}) \rb^2 \nonumber \\
&&\hspace{-0.5in}{}\times\bigg \{ f(\epsilon_{\v{p'}}) \lb 1- f(\epsilon_{\v{p}+\v{q}}) \rb \lb 1- f(\epsilon_{\v{p'}-\v{q}}) \rb \nonumber \\
&&\hspace{-0.25in}{}-\lb 1- f(\epsilon_{\v{p'}}) \rb f(\epsilon_{\v{p}+\v{q}})f(\epsilon_{\v{p'}-\v{q}})\bigg \} \label{eq:broadening-anytemp}
\ee
where $f(\epsilon)=\frac{1}{e^{\beta (\epsilon-\mu)}+1} $ is the Fermi function,   $\epsilon_{p}\equiv p^2/(2m)$, and $v(\v{q})$ is the Fourier transform of the interaction potential, here the Fourier transform of $V_{11}(\v{\rho})$.
This can be understood by the decay processes occuring in Fermi's Golden Rule, perturbing in the interactions.  For a thermal gas, this simplifies to
\be
\Gamma(\v{p},\omega) &=& \frac{1}{2(2\pi)^3}\int d\v{p'} d\v{q} \, \delta(\omega+ \epsilon_{\v{p}'} -\epsilon_{\v{p}+\v{q}} - \epsilon_{\v{p'} - \v{q}}) \nonumber \\
&&\hspace{-0.5in}{}\times\lb v(-\v{q}) - v(\v{p}-\v{p}' + \v{q}) \rb^2  e^{-\beta(\epsilon_{\v{p'}}-\mu)}. \label{eq:broadening-starting-point}
\ee
Rather than worry about the full $\v{p}$ and $\omega$ dependence of this broadening, we calculate the average broadening of the line and for each $\v{p}$ treat the frequency as being fixed to  $\omega=\epsilon_{p}$.  With this, the typical broadening --- that is, the broadening averaged over momenta $\Gamma_{\text{typ}} = (1/N)\sum_{\v{p}} \Gamma(\v{p},\epsilon_p)$ --- is
\be
\Gamma_{\text{typ}} &=& \frac{1}{n} \int \! \frac{d\v{p}}{(2\pi)^2}
\Gamma(\v{p},\epsilon_p).
\ee

Calculating $\Gamma_{\text{typ}}$ from Eq.~\eqref{eq:broadening-starting-point} requires the Fourier transform of the interaction potential.  The Fourier transform, neglecting $A_{11}$ as argued to be valid earlier, is simply
\be
v(\v{q}) &=& C_0 - 2\pi \gamma_{11} q+O(q^2),
\ee
for some constant $C_0$.  The constant $C_0$ cancels in Eq.~\eqref{eq:broadening-starting-point}.  Physically, this is because it describes a renormalization of the contact potential, which affects nothing in a gas of identical fermions.
Then, evaluating the integrals,
 the typical quasiparticle lifetime is
$ \Gamma_{\text{typ}} = \frac{m^2 (\pi+2) \gamma_{11}^2}{2\pi^3}Tn$.
Ref.~\cite{pethick:pseudopot-breakdown} argued that the recoil-free spectral broadening is simply the expression for the quasiparticle decay rate with $\gamma_{11}\rightarrow \gamma_{12}-\gamma_{11} $, giving the collisional spectral broadening
\be
\Gamma_{\text{rf}} &=& \frac{m^2 (\pi+2)(\gamma_{12}-\gamma_{11})^2}{2\pi^3}Tn.\label{eq:Gamma-dipolecollisions}
\ee
It is useful to compare this broadening to the mean dipolar spectral shift:
\be
\frac{\Gamma_{\text{rf}}}{\expec{\omega_d}} &=& \frac{\pi+2}{4 \pi^{9/2}}m^2 T (\gamma_{12}-\gamma_{11}).
\ee
Note that this ratio decreases with temperature, and  is proportional to $\gamma_{12}-\gamma_{11}$. Thus as the shifts decrease, the ratio of the broadening to the shift also decreases.  Even for  the largest experimentally realistic differences $\gamma_{12}-\gamma_{11}$, for example setting $\gamma_{11}=0$ and taking $\gamma_{12}$ to be that associated with the permanent dipole, this ratio is about 0.02.  More typical differences will be a small fraction of this ($\lsim 10\%$ or smaller).  Given the already small shifts, this broadening is negligible, perhaps at most a hundred Hertz.

So far, we have neglected the interactions with other pancakes and the spin flip terms.  Neglecting the inter-pancake interaction is reasonable since the inter-pancake interactions are somewhat smaller than the intra-pancake interactions --- by being further away by at least one lattice spacing.
 The fact that the shift \textit{vanishes} for interactions between pancakes suggests that the broadening may be even smaller than this naive argument suggests.  Since the effect of the intra-pancake interactions is quite small, it is almost certainly an excellent approximation to neglect the inter-pancake interaction induced collisional broadening.  Regarding the spin flip terms, their maximal
 is less than the maximal direct term's, and thus the spin flip term's effect is likely to be comparably small.

\subsection{Broadening from interaction potential shape}

At sufficiently high temperatures, multiple scattering channels will become relevant, and rather than merely shifting the spectral line, interactions will also broaden it. This leads to a  broadening as one leaves the cold collisional regime.  This occurs when the thermal de Broglie wavelength $\lambda_T$ becomes less than or comparable to the range of the potential.  Here, however, the temperature is sufficiently low that $\lambda_T$ is somewhat longer than the characteristic range of the potential, even for the dipolar potential.  To see this, observe that presently $\lambda_T \sim 70$nm while the dipole length is $\sim 50 $nm.  As experiments achieve colder temperatures, $\lambda_T$ will increase.

\subsection{Higher band occupation}

We have so far neglected higher band occupations.  These have various effects, including a broadening of the spectra.  Such broadening should mostly be, relative to the shifts, on the order of the relative occupation of the higher bands, perhaps $\sim 10$-$20$\% at most here. A somewhat larger contribution can come from $s$-wave collisions between molecules in different bands. Regardless, these occupations are exponentially suppressed as the temperature is lowered, and additionally it may be possible to remove them by purification techniques: e.g. higher bands tunnel more rapidly than the lowest band, and it may be possible to remove them from the lattice on a short time scale.

\section{3D lattice\label{sec:3d-lattice}}

This section considers the recoil-free spectra of molecules in a sufficiently deep 3D lattice such that tunneling is negligible.  This limit is interesting both for its simplicity and, as we show, its utility as a direct measurement tool for numerous important correlation functions, including for states created when the tunneling is significant.  Experiments on KRb molecules in 3D lattices are underway at JILA~\cite{ye:pc}.

In the infinitely deep lattice, the eigenstates of the initial Hamiltonian, prior to application of the spectral probe, are $\ket{\{n_i\}}\equiv \bigotimes_i \ket{n_i}_i$ where $\ket{n}_i$ is the Fock state with $n$ particles at site $i$.  
Then the total spectrum is the sum over sites $j$ of delta functions whose spectral shift is that for exciting particles at site $j$; in the eigenstate $\ket{\{n_i\}}$, the shift associated with exciting site $j$ is
\be
\hspace{-0.2in}\expec{\omega}_j &=& \sum_{i\ne j} \lb V_{12}(\v{R_{ij}})-V_{11}(\v{R_{ij}})+V_{\text{sf}}(\v{R_{ij}})
\rb n_i \label{cs-inhom}
\ee
where $\v{R_{ij}}$ is the separation of the $i$'th and $j$'th lattice sites. This is essentially the integrand/summand of Eq.~\eqref{eq:gen-shift-dipole}, since we are considering the shift associated with changing the rotational state at a single site $j$, projected to a 3D lattice.
Each interatomic separation between $i$ and $j$ will have its own unique shift, and this will allow us to spectroscopically measure the correlations at various distances.
Broadening can be neglected, as the only source of broadening when the tunneling is zero is the $\sim$Hz $p$-wave losses.

Any many-body wavefunction $\ket{\Psi}$ initially all in the $\alpha=1$ internal state can be expanded in this basis:
\be
\ket{\Psi} &=& \sum_{\{n_i\}} A_{\{n_i\}}\ket{\{n_i\}}.
\ee
The total spectrum is a sum of peaks with shifts given by Eq.~\eqref{cs-inhom}.  In particular, the total spectrum in the present zero-tunneling limit is
${\mc I}_{\text{tot}}(\omega)=\sum_{\{n_i\}} A_{\{n_i\}} {\mc I}_{\{n_i\}}(\omega)$ where ${\mc I}_{\{n_i\}}(\omega)$ is the spectrum for the eigenstate $\ket{\{n_i\}}$.  This spectrum is ${\mc I}_{\{n_i\}}(\omega)=\sum_j n_j \delta(\omega-\expec{\omega}_j)$ with $\expec{\omega}_j$ given by Eq.~\eqref{cs-inhom}.

For simplicity we consider the case where the 3D lattice is broken into 2D sheets such that the inter-layer interactions are negligible. This can be achieved either by depopulating all but one sheet or by increasing the lattice spacing along one axis.  The basic idea to read off correlations, discussed in the next paragraph, can be used in  more general geometries, but the procedure becomes more complicated.

\begin{figure}
\setlength{\unitlength}{1.0in}
\includegraphics[width=3.3in,angle=0]{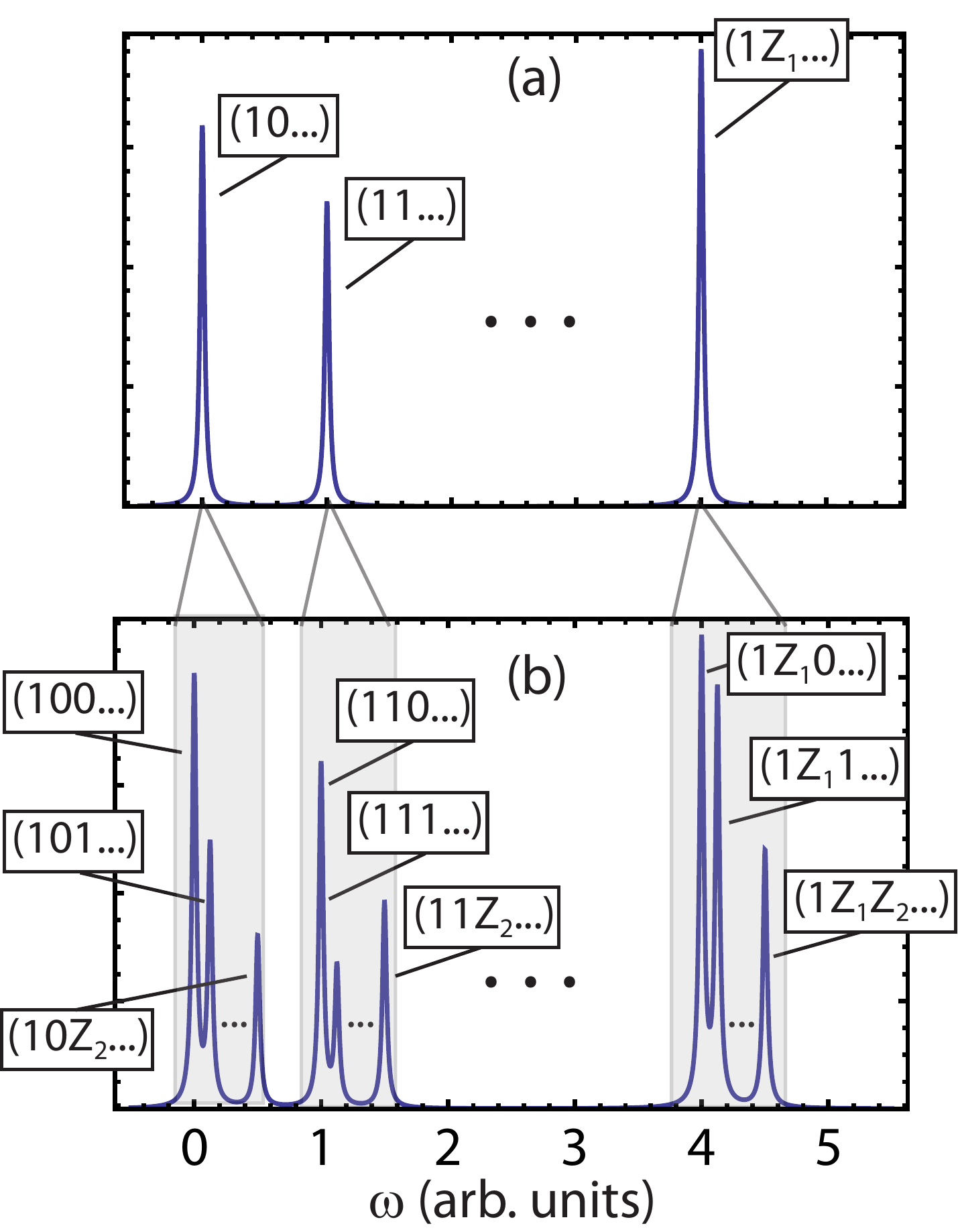}
  \caption{ \label{fig:cs-long-range}
  Zero tunneling recoil-free spectra, illustrating how to read off correlations.  (a) Recoil-free spectra with nearest neighbor interactions only.
  The correspondence between spectral shift and spatial correlations is labeled, where the peak labeled $(n_1 \, n_2 \, n_3)$ has spectral weight that is a sum over all configurations and sites $i$ with $n_1$ particles on site $i$,  $n_2$ in $i$'s nearest neighbor sites, and $n_3$ in the next nearest neighbor sites, summed over all sites $i$.  The ``$\ldots$" indicates further correlations are irrelevant to the shift, and, independent of their values at longer range, all states contribute to the peak.  Here, $Z_1$ and $Z_2$ indicate the number of nearest and next nearest neighbor sites; note the maximum number of atoms on (next) nearest neighbor sites is $Z_1$ ($Z_2$).
  (b) Recoil-free spectra with nearest and next-nearest neighbor interactions. Longer range interactions will continue to split the peaks, and with increasing spectral resolution, longer range correlations can be read off the spectrum.
  Peak heights are drawn arbitrarily.
  }
\end{figure}

Figure~\ref{fig:cs-long-range} shows an example of how to read correlation functions from the splitting of spectral peaks, explained in the caption.
 Fig.~\ref{fig:cs-long-range}~(a) shows this including only  nearest neighbor interactions, and Fig.~\ref{fig:cs-long-range}~(b) includes next-nearest neighbor interactions.  Contributions from further separated sites' correlations  will continue the pattern of splitting the lines by smaller and smaller values.

Thus we observe that recoil-free spectra can be used to cleanly extract a tremendous amount of information about the \textit{static} correlation functions of a system in a 3D lattice that have either zero or one particles per site (somewhat higher occupations can be accounted for straightforwardly, but lead to complications where  peaks start to overlap).  In particular, one can measure the joint probability distribution $P(1,n_1,n_2,\ldots)$, the probability of a site having one particle on it,  $n_1$  atoms on all the nearest neighbor sites, $n_2$ the number of atoms on all the next-nearest neighbor sites, and so on.  Neglecting occupations of more than one molecule per site, the arguments above show that this function $P(1,n_1,n_2,\ldots)$ is simply proportional to the spectral weight of a particular spectral peak.

Measuring $P(1,n_1,\ldots, n_j)$ requires finer spatial resolution as $j$ increases.  For a given string of occupancies $(n_1,\ldots,n_{j-1})$, Eq.~\eqref{cs-inhom} gives the frequency splitting between having $n_j$ and $n_j+1$ particles at lattice sites that are  $j$'th nearest neighbors to be  $\delta=V_{12}(\v{r_j})-V_{11}(\v{r_j})+V_{\text{sf}}(\v{r_{j}})$ where $\v{r_j}$ is the lattice vector corresponding to the $j$'th nearest neighbor separation.
The energy scales for KRb for current 532~nm lattice spacing square lattices (neglecting interactions between the 2D sheets that occur in a 3D lattice, and assuming that the difference in dipole moments for the initial state and final state can be made comparable to half the permanent dipole moment) enable one to resolve $P(1,n_1,n_2)$ if one has 30Hz spectral resolution and $P(1,n_1,n_2,n_3)$ if one can obtain 10Hz spectral resolution.  For other molecules, the measurements can be even more informative. For example, with RbCs one can measure $P(1,n_1,\ldots, n_4)$ using a 30Hz spectral resolution and $P(1,n_1,\ldots,n_7)$ using a  10Hz spectral resolution.  Shorter lattice spacings  enhance the range of correlations measurable for a given spectral resolutions.

Note that this joint probability distribution includes moments, e.g. $\expec{n^2}$ and $\expec{n_0n_3}$, but is substantially more informative.  In particular, one can compute \textit{any} moment between $n_j$'s that can be resolved: i.e. it gives $\expec{n_0^{\alpha_0} n_1^{\alpha_1} n_2^{\alpha_2}\ldots n_j^{\alpha_j}}$ for arbitrary $(\alpha_0,\alpha_1,\ldots, \alpha_j)$.

Importantly, this technique can probe correlations of many-body systems generated by the system with a shallower lattice, where tunneling may be important.  In this case, one can rapidly ramp up the lattice depth from the value where the physics is of interest to a value where the tunneling is effectively zero. This projects  the initial state onto a basis of the zero-tunneling eigenstates $\ket{\{n_i\}}$, and the dynamics during the interrogation then are described by the considerations above.  Thus one can measure the joint number distribution function of the initial state.  This is  analogous to the measurement of on-site number statistics done in collapse-and-revival experiments~\cite{greiner:collapse_2002,sebby-strabley:preparing_2007,sebby-strabley:preparing_2007}: a shallow lattice equilibrium state is quenched to a deep lattice state for measurement.

Such a technique is straightforwardly extended to measure correlations where the initial state may have several rotational levels populated, and to measure inter-site correlation functions between atoms in these various states. This occurs naturally in, and would be particularly useful for, proposals to use molecules' internal degrees of freedom as ``pseudospins" to emulate quantum magnetism~\cite{gorshkov:dipoles-tjvw-short,gorshkov:dipoles-tjvw-long}, where the technique would probe the pseudospin correlations.

Here we explicitly considered interpreting these spectra for the 2D lattice in a 2D system.  This is straightforwardly extended to a 3D lattice, but the form of the $V_{12} - V_{11}+V_{\text{sf}}$ will be slightly more complicated due to interaction anisotropies.  However, our discussion doesn't rely crucially on the form of $V_{12}-V_{11}+V_{\text{sf}}$, except for the fact it is long ranged.

Finally, this measurement technique is not limited to molecules. For example, it may be possible to measure correlations of alkali atoms in optical lattices by exciting all the atoms to a Rydberg state immediately before the spectral interrogation.  The spectral pulse would then chosen to drive the atoms between two Rydberg states, which will experience  dipolar effects in a manner analogous to the molecules.

\section{Summary}

In this paper, we have calculated the recoil-free spectra for dipolar molecules in one dimensional lattices of two-dimensional pancakes, and in deep three dimensional lattices.  We particularly focused on the case of driving transitions between different rotational states. We have incorporated both the dipolar interactions and the $p$-wave interactions and losses, the latter of which is important for reactive molecules. We gave  general expressions for the shifts, and for the weakly interacting gas evaluated these expressions, as well as their simpler low temperature ($T\ll \mu$, degenerate) and high temperature ($T\gg \mu$, thermal) limits. We calculated the broadening of these lines from collisions, losses, differing polarizabilities of rotational states, and other mechanisms.
We overviewed typical scales for these various effects for the case of KRb molecules.  The overall scale for the trap summed spectrum under current experimental conditions is on the order of $\sim 5$kHz for strong dipoles due to the dipolar interactions and the inhomogeneous density of the trapped system. This will increase to a constant with decreasing temperature and increase indefinitely with increasing density (until perturbation theory breaks down). Observation of this lineshape will confirm that the interactions in this system take the expected form and can be used to diagnose temperature and  degeneracy. Deviations from the weakly-interacting spectral lineshapes can then be used to observe and characterize strong correlation physics.

The $p$-wave losses induce a few Hertz broadening, and the real $p$-wave interactions should be very small.  Collisional broadening is also small, probably much less than 100Hz for current conditions.  The largest source of shift and broadening, other than the dipolar interactions, comes from the differing  polarizabilities of rotational states. This is on the order of a few kilohertz, and decreases with temperature. However it can be eliminated by working, for example, at the ``magic angle" or ``magic field strength"~\cite{kotochigova:electric_2010}.  Recoil-free spectra offer a useful probe of this physics.

In the three-dimensional lattice, we showed how the spectra give access to a multitude of static correlation functions that are inaccessible by other techniques.  The spectral weight in each peak is proportional to the probability of a certain set of spatial configurations of a particular set of lattice occupations.  This allows one to measure joint probability distributions for correlations among a few nearest neighbor sites.

We mention that although we focused on the most immediately relevant case of dipoles aligned perpendicular  to the pancakes and in a one dimensional lattice of two dimensional pancakes, most of our general formulas extend straightforwardly to other cases, for example, three dimensions or other angles of alignment. Only an evaluation  of the integrals changes.  Presumably, many of the overall magnitudes typically will remain similar.

Finally, although we focused on dipolar molecules in our quantitative estimates, the effects are relevant elsewhere.  Rydberg atoms have a strong dipolar interaction. In alkaline earth atomic clocks, $p$-wave interactions and reactive losses can be important, and although the dipolar interactions are very small, they may be important as clock accuracies approach their fundamental limits.
However,  we mention that we have treated the dipole interaction electrostatically. This is valid when the particle separation is much less than the wavelength of the photons mediating the dipole-dipole interaction.  For molecules we expect this to be quantitatively accurate.  However, in other contexts such as atomic clocks, the dynamic field becomes important when interparticle separations are comparable to or larger than the wavelength of the photon associated with the dipole transition~\cite{chang:controlling_2004}.  In this case, the treatment of dipolar interactions of the excited state is quite different.  For example, they scale as $1/r$ where $r$ is the interparticle separation.  In this context, careful choice of lattice parameters
mitigates these effects~\cite{chang:controlling_2004}.

\textit{Acknowledgements}.---We thank Jun Ye, Debbie Jin, Salvatore Manmana, Michael Foss-Feig, Erich Mueller, John Bohn, Goulven Qu{\'e}m{\'e}ner, and the molecule experiment group at JILA for numerous discussions. KRAH and AMR were supported by grants from the NSF (PFC and PIF-
0904017), the AFOSR, and a grant from the ARO with
funding from the DARPA-OLE.  AVG was supported by the Lee A. DuBridge Fellowship and a grant from the NSF (PHY-0803371).

\bibliography{molecule-rf}

\begin{thebibliography}{59}
\expandafter\ifx\csname natexlab\endcsname\relax\def\natexlab#1{#1}\fi
\expandafter\ifx\csname bibnamefont\endcsname\relax
  \def\bibnamefont#1{#1}\fi
\expandafter\ifx\csname bibfnamefont\endcsname\relax
  \def\bibfnamefont#1{#1}\fi
\expandafter\ifx\csname citenamefont\endcsname\relax
  \def\citenamefont#1{#1}\fi
\expandafter\ifx\csname url\endcsname\relax
  \def\url#1{\texttt{#1}}\fi
\expandafter\ifx\csname urlprefix\endcsname\relax\def\urlprefix{URL }\fi
\providecommand{\bibinfo}[2]{#2}
\providecommand{\eprint}[2][]{\url{#2}}

\bibitem[{\citenamefont{Fried et~al.}(1998)\citenamefont{Fried, Killian,
  Willmann, Landhuis, Moss, Kleppner, and Greytak}}]{fried:hydrogen}
\bibinfo{author}{\bibfnamefont{D.~G.} \bibnamefont{Fried}},
  \bibinfo{author}{\bibfnamefont{T.~C.} \bibnamefont{Killian}},
  \bibinfo{author}{\bibfnamefont{L.}~\bibnamefont{Willmann}},
  \bibinfo{author}{\bibfnamefont{D.}~\bibnamefont{Landhuis}},
  \bibinfo{author}{\bibfnamefont{S.~C.} \bibnamefont{Moss}},
  \bibinfo{author}{\bibfnamefont{D.}~\bibnamefont{Kleppner}}, \bibnamefont{and}
  \bibinfo{author}{\bibfnamefont{T.~J.} \bibnamefont{Greytak}},
  \bibinfo{journal}{Phys. Rev. Lett.} \textbf{\bibinfo{volume}{81}},
  \bibinfo{pages}{3811} (\bibinfo{year}{1998}).

\bibitem[{\citenamefont{Campbell et~al.}(2006)\citenamefont{Campbell, Mun,
  Boyd, Medley, Leanhardt, Marcassa, Pritchard, and
  Ketterle}}]{campbell_imagingmott_2006-2}
\bibinfo{author}{\bibfnamefont{G.~K.} \bibnamefont{Campbell}},
  \bibinfo{author}{\bibfnamefont{J.}~\bibnamefont{Mun}},
  \bibinfo{author}{\bibfnamefont{M.}~\bibnamefont{Boyd}},
  \bibinfo{author}{\bibfnamefont{P.}~\bibnamefont{Medley}},
  \bibinfo{author}{\bibfnamefont{A.~E.} \bibnamefont{Leanhardt}},
  \bibinfo{author}{\bibfnamefont{L.~G.} \bibnamefont{Marcassa}},
  \bibinfo{author}{\bibfnamefont{D.~E.} \bibnamefont{Pritchard}},
  \bibnamefont{and} \bibinfo{author}{\bibfnamefont{W.}~\bibnamefont{Ketterle}},
  \bibinfo{journal}{Science} \textbf{\bibinfo{volume}{313}},
  \bibinfo{pages}{649} (\bibinfo{year}{2006}).

\bibitem[{\citenamefont{Ohashi et~al.}(2006)\citenamefont{Ohashi, Kitaura, and
  Matsumoto}}]{ohashi_itinerant-localized_2006}
\bibinfo{author}{\bibfnamefont{Y.}~\bibnamefont{Ohashi}},
  \bibinfo{author}{\bibfnamefont{M.}~\bibnamefont{Kitaura}}, \bibnamefont{and}
  \bibinfo{author}{\bibfnamefont{H.}~\bibnamefont{Matsumoto}},
  \bibinfo{journal}{Phys. Rev. A} \textbf{\bibinfo{volume}{73}},
  \bibinfo{pages}{033617} (\bibinfo{year}{2006}).

\bibitem[{\citenamefont{Hazzard and Mueller}(2007)}]{hazzard_hyperfine_2007}
\bibinfo{author}{\bibfnamefont{K.~R.~A.} \bibnamefont{Hazzard}}
  \bibnamefont{and} \bibinfo{author}{\bibfnamefont{E.~J.}
  \bibnamefont{Mueller}}, \bibinfo{journal}{Phys. Rev. A}
  \textbf{\bibinfo{volume}{76}} (\bibinfo{year}{2007}).

\bibitem[{\citenamefont{Hazzard and
  Mueller}(2010{\natexlab{a}})}]{hazzard_many-body_2009}
\bibinfo{author}{\bibfnamefont{K.~R.~A.} \bibnamefont{Hazzard}}
  \bibnamefont{and} \bibinfo{author}{\bibfnamefont{E.~J.}
  \bibnamefont{Mueller}}, \bibinfo{journal}{Phys. Rev. A}
  \textbf{\bibinfo{volume}{81}}, \bibinfo{pages}{033404}
  (\bibinfo{year}{2010}{\natexlab{a}}).

\bibitem[{\citenamefont{Sun et~al.}(2009)\citenamefont{Sun, Lannert, and
  Vishveshwara}}]{sun_probing_2009}
\bibinfo{author}{\bibfnamefont{K.}~\bibnamefont{Sun}},
  \bibinfo{author}{\bibfnamefont{C.}~\bibnamefont{Lannert}}, \bibnamefont{and}
  \bibinfo{author}{\bibfnamefont{S.}~\bibnamefont{Vishveshwara}},
  \bibinfo{journal}{Phys. Rev. A} \textbf{\bibinfo{volume}{79}},
  \bibinfo{pages}{043422} (\bibinfo{year}{2009}).

\bibitem[{\citenamefont{Schunck et~al.}(2008)\citenamefont{Schunck, il~Shin,
  Schirotzek, and Ketterle}}]{schunck:determination_2008}
\bibinfo{author}{\bibfnamefont{C.~H.} \bibnamefont{Schunck}},
  \bibinfo{author}{\bibfnamefont{Y.}~\bibnamefont{il~Shin}},
  \bibinfo{author}{\bibfnamefont{A.}~\bibnamefont{Schirotzek}},
  \bibnamefont{and} \bibinfo{author}{\bibfnamefont{W.}~\bibnamefont{Ketterle}},
  \bibinfo{journal}{Nature} \textbf{\bibinfo{volume}{454}},
  \bibinfo{pages}{793} (\bibinfo{year}{2008}).

\bibitem[{\citenamefont{Schirotzek et~al.}(2008)\citenamefont{Schirotzek, Shin,
  Schunck, and Ketterle}}]{schirotzek:determination_2008}
\bibinfo{author}{\bibfnamefont{A.}~\bibnamefont{Schirotzek}},
  \bibinfo{author}{\bibfnamefont{Y.-i.} \bibnamefont{Shin}},
  \bibinfo{author}{\bibfnamefont{C.~H.} \bibnamefont{Schunck}},
  \bibnamefont{and} \bibinfo{author}{\bibfnamefont{W.}~\bibnamefont{Ketterle}},
  \bibinfo{journal}{Phys. Rev. Lett.} \textbf{\bibinfo{volume}{101}},
  \bibinfo{pages}{140403} (\bibinfo{year}{2008}).

\bibitem[{\citenamefont{Schirotzek et~al.}(2009)\citenamefont{Schirotzek, Wu,
  Sommer, and Zwierlein}}]{schirotzek:observation_2009}
\bibinfo{author}{\bibfnamefont{A.}~\bibnamefont{Schirotzek}},
  \bibinfo{author}{\bibfnamefont{C.-H.} \bibnamefont{Wu}},
  \bibinfo{author}{\bibfnamefont{A.}~\bibnamefont{Sommer}}, \bibnamefont{and}
  \bibinfo{author}{\bibfnamefont{M.~W.} \bibnamefont{Zwierlein}},
  \bibinfo{journal}{Phys. Rev. Lett.} \textbf{\bibinfo{volume}{102}},
  \bibinfo{pages}{230402} (\bibinfo{year}{2009}).

\bibitem[{\citenamefont{Stewart et~al.}(2008)\citenamefont{Stewart, Gaebler,
  and Jin}}]{stewart:using_2008}
\bibinfo{author}{\bibfnamefont{J.~T.} \bibnamefont{Stewart}},
  \bibinfo{author}{\bibfnamefont{J.~P.} \bibnamefont{Gaebler}},
  \bibnamefont{and} \bibinfo{author}{\bibfnamefont{D.~S.} \bibnamefont{Jin}},
  \bibinfo{journal}{Nature} \textbf{\bibinfo{volume}{454}},
  \bibinfo{pages}{744} (\bibinfo{year}{2008}).

\bibitem[{\citenamefont{Gaebler et~al.}(2010)\citenamefont{Gaebler, Stewart,
  Drake, Jin, Perali, Pieri, and Strinati}}]{gaebler:observation_2010}
\bibinfo{author}{\bibfnamefont{J.~P.} \bibnamefont{Gaebler}},
  \bibinfo{author}{\bibfnamefont{J.~T.} \bibnamefont{Stewart}},
  \bibinfo{author}{\bibfnamefont{T.~E.} \bibnamefont{Drake}},
  \bibinfo{author}{\bibfnamefont{D.~S.} \bibnamefont{Jin}},
  \bibinfo{author}{\bibfnamefont{A.}~\bibnamefont{Perali}},
  \bibinfo{author}{\bibfnamefont{P.}~\bibnamefont{Pieri}}, \bibnamefont{and}
  \bibinfo{author}{\bibfnamefont{G.~C.} \bibnamefont{Strinati}},
  \bibinfo{journal}{Nature Physics} \textbf{\bibinfo{volume}{6}},
  \bibinfo{pages}{569} (\bibinfo{year}{2010}).

\bibitem[{\citenamefont{Perali et~al.}(2011)\citenamefont{Perali, Palestini,
  Pieri, Strinati, Stewart, Gaebler, Drake, and Jin}}]{perali:evolution_2011}
\bibinfo{author}{\bibfnamefont{A.}~\bibnamefont{Perali}},
  \bibinfo{author}{\bibfnamefont{F.}~\bibnamefont{Palestini}},
  \bibinfo{author}{\bibfnamefont{P.}~\bibnamefont{Pieri}},
  \bibinfo{author}{\bibfnamefont{G.~C.} \bibnamefont{Strinati}},
  \bibinfo{author}{\bibfnamefont{J.~T.} \bibnamefont{Stewart}},
  \bibinfo{author}{\bibfnamefont{J.~P.} \bibnamefont{Gaebler}},
  \bibinfo{author}{\bibfnamefont{T.~E.} \bibnamefont{Drake}}, \bibnamefont{and}
  \bibinfo{author}{\bibfnamefont{D.~S.} \bibnamefont{Jin}},
  \bibinfo{journal}{Phys. Rev. Lett.} \textbf{\bibinfo{volume}{106}},
  \bibinfo{pages}{060402} (\bibinfo{year}{2011}).

\bibitem[{\citenamefont{Lemke et~al.}(2009)\citenamefont{Lemke, Ludlow, Barber,
  Fortier, Diddams, Jiang, Jefferts, Heavner, Parker, and
  Oates}}]{lemke:spin_2009}
\bibinfo{author}{\bibfnamefont{N.~D.} \bibnamefont{Lemke}},
  \bibinfo{author}{\bibfnamefont{A.~D.} \bibnamefont{Ludlow}},
  \bibinfo{author}{\bibfnamefont{Z.~W.} \bibnamefont{Barber}},
  \bibinfo{author}{\bibfnamefont{T.~M.} \bibnamefont{Fortier}},
  \bibinfo{author}{\bibfnamefont{S.~A.} \bibnamefont{Diddams}},
  \bibinfo{author}{\bibfnamefont{Y.}~\bibnamefont{Jiang}},
  \bibinfo{author}{\bibfnamefont{S.~R.} \bibnamefont{Jefferts}},
  \bibinfo{author}{\bibfnamefont{T.~P.} \bibnamefont{Heavner}},
  \bibinfo{author}{\bibfnamefont{T.~E.} \bibnamefont{Parker}},
  \bibnamefont{and} \bibinfo{author}{\bibfnamefont{C.~W.} \bibnamefont{Oates}},
  \bibinfo{journal}{Phys. Rev. Lett.} \textbf{\bibinfo{volume}{103}},
  \bibinfo{pages}{063001} (\bibinfo{year}{2009}).

\bibitem[{\citenamefont{Swallows et~al.}(2011)\citenamefont{Swallows, Bishof,
  Lin, Blatt, Martin, Rey, and Ye}}]{swallows:suppression_2011}
\bibinfo{author}{\bibfnamefont{M.~D.} \bibnamefont{Swallows}},
  \bibinfo{author}{\bibfnamefont{M.}~\bibnamefont{Bishof}},
  \bibinfo{author}{\bibfnamefont{Y.}~\bibnamefont{Lin}},
  \bibinfo{author}{\bibfnamefont{S.}~\bibnamefont{Blatt}},
  \bibinfo{author}{\bibfnamefont{M.~J.} \bibnamefont{Martin}},
  \bibinfo{author}{\bibfnamefont{A.~M.} \bibnamefont{Rey}}, \bibnamefont{and}
  \bibinfo{author}{\bibfnamefont{J.}~\bibnamefont{Ye}},
  \bibinfo{journal}{Science} \textbf{\bibinfo{volume}{331}},
  \bibinfo{pages}{1043} (\bibinfo{year}{2011}).

\bibitem[{\citenamefont{de~Miranda et~al.}(2011)\citenamefont{de~Miranda,
  Chotia, Neyenhuis, Wang, Qu{\'e}m{\'e}ner, Ospelkaus, Bohn, Ye, and
  Jin}}]{miranda:controlling_2011}
\bibinfo{author}{\bibfnamefont{M.~H.~G.} \bibnamefont{de~Miranda}},
  \bibinfo{author}{\bibfnamefont{A.}~\bibnamefont{Chotia}},
  \bibinfo{author}{\bibfnamefont{B.}~\bibnamefont{Neyenhuis}},
  \bibinfo{author}{\bibfnamefont{D.}~\bibnamefont{Wang}},
  \bibinfo{author}{\bibfnamefont{G.}~\bibnamefont{Qu{\'e}m{\'e}ner}},
  \bibinfo{author}{\bibfnamefont{S.}~\bibnamefont{Ospelkaus}},
  \bibinfo{author}{\bibfnamefont{J.~L.} \bibnamefont{Bohn}},
  \bibinfo{author}{\bibfnamefont{J.}~\bibnamefont{Ye}}, \bibnamefont{and}
  \bibinfo{author}{\bibfnamefont{D.~S.} \bibnamefont{Jin}},
  \bibinfo{journal}{Nature Physics} \textbf{\bibinfo{volume}{7}},
  \bibinfo{pages}{502} (\bibinfo{year}{2011}).

\bibitem[{\citenamefont{Ospelkaus et~al.}(2010)\citenamefont{Ospelkaus, Ni,
  Wang, de~Miranda, Neyenhuis, Qu{\'e}m{\'e}ner, Julienne, Bohn, Jin, and
  Ye}}]{ospelkaus:quantum-state_2010}
\bibinfo{author}{\bibfnamefont{S.}~\bibnamefont{Ospelkaus}},
  \bibinfo{author}{\bibfnamefont{K.-K.} \bibnamefont{Ni}},
  \bibinfo{author}{\bibfnamefont{D.}~\bibnamefont{Wang}},
  \bibinfo{author}{\bibfnamefont{M.~H.~G.} \bibnamefont{de~Miranda}},
  \bibinfo{author}{\bibfnamefont{B.}~\bibnamefont{Neyenhuis}},
  \bibinfo{author}{\bibfnamefont{G.}~\bibnamefont{Qu{\'e}m{\'e}ner}},
  \bibinfo{author}{\bibfnamefont{P.~S.} \bibnamefont{Julienne}},
  \bibinfo{author}{\bibfnamefont{J.~L.} \bibnamefont{Bohn}},
  \bibinfo{author}{\bibfnamefont{D.~S.} \bibnamefont{Jin}}, \bibnamefont{and}
  \bibinfo{author}{\bibfnamefont{J.}~\bibnamefont{Ye}},
  \bibinfo{journal}{Science} \textbf{\bibinfo{volume}{327}},
  \bibinfo{pages}{853} (\bibinfo{year}{2010}).

\bibitem[{\citenamefont{Ni et~al.}(2008)\citenamefont{Ni, Ospelkaus,
  de~Miranda, Pe'er, Neyenhuis, Zirbel, Kotochigova, Julienne, Jin, and
  Ye}}]{ni:a_2008}
\bibinfo{author}{\bibfnamefont{K.-K.} \bibnamefont{Ni}},
  \bibinfo{author}{\bibfnamefont{S.}~\bibnamefont{Ospelkaus}},
  \bibinfo{author}{\bibfnamefont{M.~H.~G.} \bibnamefont{de~Miranda}},
  \bibinfo{author}{\bibfnamefont{A.}~\bibnamefont{Pe'er}},
  \bibinfo{author}{\bibfnamefont{B.}~\bibnamefont{Neyenhuis}},
  \bibinfo{author}{\bibfnamefont{J.~J.} \bibnamefont{Zirbel}},
  \bibinfo{author}{\bibfnamefont{S.}~\bibnamefont{Kotochigova}},
  \bibinfo{author}{\bibfnamefont{P.~S.} \bibnamefont{Julienne}},
  \bibinfo{author}{\bibfnamefont{D.~S.} \bibnamefont{Jin}}, \bibnamefont{and}
  \bibinfo{author}{\bibfnamefont{J.}~\bibnamefont{Ye}},
  \bibinfo{journal}{Science} \textbf{\bibinfo{volume}{322}},
  \bibinfo{pages}{231} (\bibinfo{year}{2008}).

\bibitem[{\citenamefont{Ni et~al.}(2010)\citenamefont{Ni, Ospelkaus, Wang,
  Qu{\'e}m{\'e}ner, Neyenhuis, de~Miranda, Bohn, Ye, and
  Jin}}]{ni:dipolar_2010}
\bibinfo{author}{\bibfnamefont{K.-K.} \bibnamefont{Ni}},
  \bibinfo{author}{\bibfnamefont{S.}~\bibnamefont{Ospelkaus}},
  \bibinfo{author}{\bibfnamefont{D.}~\bibnamefont{Wang}},
  \bibinfo{author}{\bibfnamefont{G.}~\bibnamefont{Qu{\'e}m{\'e}ner}},
  \bibinfo{author}{\bibfnamefont{B.}~\bibnamefont{Neyenhuis}},
  \bibinfo{author}{\bibfnamefont{M.~H.~G.} \bibnamefont{de~Miranda}},
  \bibinfo{author}{\bibfnamefont{J.~L.} \bibnamefont{Bohn}},
  \bibinfo{author}{\bibfnamefont{J.}~\bibnamefont{Ye}}, \bibnamefont{and}
  \bibinfo{author}{\bibfnamefont{D.~S.} \bibnamefont{Jin}},
  \bibinfo{journal}{Nature} \textbf{\bibinfo{volume}{464}},
  \bibinfo{pages}{1324} (\bibinfo{year}{2010}).

\bibitem[{\citenamefont{Ospelkaus et~al.}(2009)\citenamefont{Ospelkaus, Ni,
  de~Miranda, Neyenhuis, Wang, Kotochigova, Julienne, Jin, and
  Ye}}]{ospelkaus:ultracold_2009}
\bibinfo{author}{\bibfnamefont{S.}~\bibnamefont{Ospelkaus}},
  \bibinfo{author}{\bibfnamefont{K.-K.} \bibnamefont{Ni}},
  \bibinfo{author}{\bibfnamefont{M.~H.~G.} \bibnamefont{de~Miranda}},
  \bibinfo{author}{\bibfnamefont{B.}~\bibnamefont{Neyenhuis}},
  \bibinfo{author}{\bibfnamefont{D.}~\bibnamefont{Wang}},
  \bibinfo{author}{\bibfnamefont{S.}~\bibnamefont{Kotochigova}},
  \bibinfo{author}{\bibfnamefont{P.~S.} \bibnamefont{Julienne}},
  \bibinfo{author}{\bibfnamefont{D.~S.} \bibnamefont{Jin}}, \bibnamefont{and}
  \bibinfo{author}{\bibfnamefont{J.}~\bibnamefont{Ye}},
  \bibinfo{journal}{Faraday Discussions} \textbf{\bibinfo{volume}{142}},
  \bibinfo{pages}{351} (\bibinfo{year}{2009}).

\bibitem[{\citenamefont{Wang et~al.}(2010)\citenamefont{Wang, Neyenhuis,
  de~Miranda, Ni, Ospelkaus, Jin, and Ye}}]{wang_direct_2010}
\bibinfo{author}{\bibfnamefont{D.}~\bibnamefont{Wang}},
  \bibinfo{author}{\bibfnamefont{B.}~\bibnamefont{Neyenhuis}},
  \bibinfo{author}{\bibfnamefont{M.~H.~G.} \bibnamefont{de~Miranda}},
  \bibinfo{author}{\bibfnamefont{K.-K.} \bibnamefont{Ni}},
  \bibinfo{author}{\bibfnamefont{S.}~\bibnamefont{Ospelkaus}},
  \bibinfo{author}{\bibfnamefont{D.~S.} \bibnamefont{Jin}}, \bibnamefont{and}
  \bibinfo{author}{\bibfnamefont{J.}~\bibnamefont{Ye}}, \bibinfo{journal}{Phys.
  Rev. A} \textbf{\bibinfo{volume}{81}}, \bibinfo{pages}{061404}
  (\bibinfo{year}{2010}).

\bibitem[{\citenamefont{Lahaye et~al.}(2009)\citenamefont{Lahaye, Menotti,
  Santos, Lewenstein, and Pfau}}]{lahaye_physics_2009}
\bibinfo{author}{\bibfnamefont{T.}~\bibnamefont{Lahaye}},
  \bibinfo{author}{\bibfnamefont{C.}~\bibnamefont{Menotti}},
  \bibinfo{author}{\bibfnamefont{L.}~\bibnamefont{Santos}},
  \bibinfo{author}{\bibfnamefont{M.}~\bibnamefont{Lewenstein}},
  \bibnamefont{and} \bibinfo{author}{\bibfnamefont{T.}~\bibnamefont{Pfau}},
  \bibinfo{journal}{Rep. Prog. Phys.} \textbf{\bibinfo{volume}{72}},
  \bibinfo{pages}{126401} (\bibinfo{year}{2009}), ISSN
  \bibinfo{issn}{0034-4885}.

\bibitem[{\citenamefont{Pupillo et~al.}(2008)\citenamefont{Pupillo, Micheli,
  B{\"u}chler, and Zoller}}]{pupillo:condensed_2008}
\bibinfo{author}{\bibfnamefont{G.}~\bibnamefont{Pupillo}},
  \bibinfo{author}{\bibfnamefont{A.}~\bibnamefont{Micheli}},
  \bibinfo{author}{\bibfnamefont{H.}~\bibnamefont{B{\"u}chler}},
  \bibnamefont{and} \bibinfo{author}{\bibfnamefont{P.}~\bibnamefont{Zoller}},
  \emph{\bibinfo{title}{Cold molecules: Creation and applications}}
  (\bibinfo{publisher}{Taylor \& Francis}, \bibinfo{year}{2008}), chap.
  \bibinfo{chapter}{Condensed Matter Physics with Cold Polar Molecules}.

\bibitem[{\citenamefont{Baranov}(2008)}]{baranov:theoretical_2008}
\bibinfo{author}{\bibfnamefont{M.}~\bibnamefont{Baranov}},
  \bibinfo{journal}{Physics Reports} \textbf{\bibinfo{volume}{464}},
  \bibinfo{pages}{71 } (\bibinfo{year}{2008}), ISSN \bibinfo{issn}{0370-1573}.

\bibitem[{\citenamefont{Carr et~al.}(2009)\citenamefont{Carr, DeMille, Krems,
  and Ye}}]{carr:cold_2009}
\bibinfo{author}{\bibfnamefont{L.~D.} \bibnamefont{Carr}},
  \bibinfo{author}{\bibfnamefont{D.}~\bibnamefont{DeMille}},
  \bibinfo{author}{\bibfnamefont{R.~V.} \bibnamefont{Krems}}, \bibnamefont{and}
  \bibinfo{author}{\bibfnamefont{J.}~\bibnamefont{Ye}}, \bibinfo{journal}{New
  Journal of Physics} \textbf{\bibinfo{volume}{11}}, \bibinfo{pages}{055049}
  (\bibinfo{year}{2009}).

\bibitem[{\citenamefont{Lin et~al.}(2010)\citenamefont{Lin, Zhao, and
  Liu}}]{lin:liquid_2010}
\bibinfo{author}{\bibfnamefont{C.}~\bibnamefont{Lin}},
  \bibinfo{author}{\bibfnamefont{E.}~\bibnamefont{Zhao}}, \bibnamefont{and}
  \bibinfo{author}{\bibfnamefont{W.~V.} \bibnamefont{Liu}},
  \bibinfo{journal}{Phys. Rev. B} \textbf{\bibinfo{volume}{81}},
  \bibinfo{pages}{045115} (\bibinfo{year}{2010}).

\bibitem[{\citenamefont{Quintanilla et~al.}(2009)\citenamefont{Quintanilla,
  Carr, and Betouras}}]{quintanilla:metanematic_2009}
\bibinfo{author}{\bibfnamefont{J.}~\bibnamefont{Quintanilla}},
  \bibinfo{author}{\bibfnamefont{S.~T.} \bibnamefont{Carr}}, \bibnamefont{and}
  \bibinfo{author}{\bibfnamefont{J.~J.} \bibnamefont{Betouras}},
  \bibinfo{journal}{Phys. Rev. A} \textbf{\bibinfo{volume}{79}},
  \bibinfo{pages}{031601} (\bibinfo{year}{2009}).

\bibitem[{\citenamefont{Sun et~al.}(2010)\citenamefont{Sun, Wu, and
  Das~Sarma}}]{sun:spontaneous_2010}
\bibinfo{author}{\bibfnamefont{K.}~\bibnamefont{Sun}},
  \bibinfo{author}{\bibfnamefont{C.}~\bibnamefont{Wu}}, \bibnamefont{and}
  \bibinfo{author}{\bibfnamefont{S.}~\bibnamefont{Das~Sarma}},
  \bibinfo{journal}{Phys. Rev. B} \textbf{\bibinfo{volume}{82}},
  \bibinfo{pages}{075105} (\bibinfo{year}{2010}).

\bibitem[{\citenamefont{Cheng}(2009)}]{cheng:static_2009}
\bibinfo{author}{\bibfnamefont{S.-C.} \bibnamefont{Cheng}},
  \bibinfo{journal}{arxiv:0902.2276}  (\bibinfo{year}{2009}).

\bibitem[{\citenamefont{Baranov et~al.}(2008)\citenamefont{Baranov, Fehrmann,
  and Lewenstein}}]{baranov:wigner_2008}
\bibinfo{author}{\bibfnamefont{M.~A.} \bibnamefont{Baranov}},
  \bibinfo{author}{\bibfnamefont{H.}~\bibnamefont{Fehrmann}}, \bibnamefont{and}
  \bibinfo{author}{\bibfnamefont{M.}~\bibnamefont{Lewenstein}},
  \bibinfo{journal}{Phys. Rev. Lett.} \textbf{\bibinfo{volume}{100}},
  \bibinfo{pages}{200402} (\bibinfo{year}{2008}).

\bibitem[{\citenamefont{Baranov et~al.}(2010)\citenamefont{Baranov, Micheli,
  Ronen, and Zoller}}]{baranov:bilayer_2010}
\bibinfo{author}{\bibfnamefont{M.}~\bibnamefont{Baranov}},
  \bibinfo{author}{\bibfnamefont{A.}~\bibnamefont{Micheli}},
  \bibinfo{author}{\bibfnamefont{S.}~\bibnamefont{Ronen}}, \bibnamefont{and}
  \bibinfo{author}{\bibfnamefont{P.}~\bibnamefont{Zoller}}, p.
  \bibinfo{pages}{arXiv:1012.5589} (\bibinfo{year}{2010}).

\bibitem[{\citenamefont{Bruun and Taylor}(2008)}]{bruun:quantum_2010}
\bibinfo{author}{\bibfnamefont{G.~M.} \bibnamefont{Bruun}} \bibnamefont{and}
  \bibinfo{author}{\bibfnamefont{E.}~\bibnamefont{Taylor}},
  \bibinfo{journal}{Phys. Rev. Lett.} \textbf{\bibinfo{volume}{101}},
  \bibinfo{pages}{245301} (\bibinfo{year}{2008}).

\bibitem[{\citenamefont{Wang et~al.}(2006)\citenamefont{Wang, Lukin, and
  Demler}}]{wang:quantum_2006}
\bibinfo{author}{\bibfnamefont{D.-W.} \bibnamefont{Wang}},
  \bibinfo{author}{\bibfnamefont{M.~D.} \bibnamefont{Lukin}}, \bibnamefont{and}
  \bibinfo{author}{\bibfnamefont{E.}~\bibnamefont{Demler}},
  \bibinfo{journal}{Phys. Rev. Lett.} \textbf{\bibinfo{volume}{97}},
  \bibinfo{pages}{180413} (\bibinfo{year}{2006}).

\bibitem[{\citenamefont{Trefzger et~al.}(2011)\citenamefont{Trefzger, Menotti,
  Capogrosso-Sansone, and Lewenstein}}]{trefzger:ultracold_2011}
\bibinfo{author}{\bibfnamefont{C.}~\bibnamefont{Trefzger}},
  \bibinfo{author}{\bibfnamefont{C.}~\bibnamefont{Menotti}},
  \bibinfo{author}{\bibfnamefont{B.}~\bibnamefont{Capogrosso-Sansone}},
  \bibnamefont{and}
  \bibinfo{author}{\bibfnamefont{M.}~\bibnamefont{Lewenstein}}, p.
  \bibinfo{pages}{arxiv:1103.3145} (\bibinfo{year}{2011}).

\bibitem[{\citenamefont{Osterloh et~al.}(2007)\citenamefont{Osterloh,
  Barber\'an, and Lewenstein}}]{osterloh:strongly_2007}
\bibinfo{author}{\bibfnamefont{K.}~\bibnamefont{Osterloh}},
  \bibinfo{author}{\bibfnamefont{N.}~\bibnamefont{Barber\'an}},
  \bibnamefont{and}
  \bibinfo{author}{\bibfnamefont{M.}~\bibnamefont{Lewenstein}},
  \bibinfo{journal}{Phys. Rev. Lett.} \textbf{\bibinfo{volume}{99}},
  \bibinfo{pages}{160403} (\bibinfo{year}{2007}).

\bibitem[{\citenamefont{Qiu et~al.}(2011)\citenamefont{Qiu, Kou, Hu, Wan, and
  Yi}}]{qiu:quantum_2011}
\bibinfo{author}{\bibfnamefont{R.-Z.} \bibnamefont{Qiu}},
  \bibinfo{author}{\bibfnamefont{S.-P.} \bibnamefont{Kou}},
  \bibinfo{author}{\bibfnamefont{Z.-X.} \bibnamefont{Hu}},
  \bibinfo{author}{\bibfnamefont{X.}~\bibnamefont{Wan}}, \bibnamefont{and}
  \bibinfo{author}{\bibfnamefont{S.}~\bibnamefont{Yi}}, p.
  \bibinfo{pages}{arXiv:1104.0100} (\bibinfo{year}{2011}).

\bibitem[{\citenamefont{Cheng}(2008)}]{cheng:excitation_2008}
\bibinfo{author}{\bibfnamefont{S.-C.} \bibnamefont{Cheng}},
  \bibinfo{journal}{arXiv:0809.4972}  (\bibinfo{year}{2008}).

\bibitem[{\citenamefont{Baranov et~al.}(2005)\citenamefont{Baranov, Osterloh,
  and Lewenstein}}]{baranov:fractional_2005}
\bibinfo{author}{\bibfnamefont{M.~A.} \bibnamefont{Baranov}},
  \bibinfo{author}{\bibfnamefont{K.}~\bibnamefont{Osterloh}}, \bibnamefont{and}
  \bibinfo{author}{\bibfnamefont{M.}~\bibnamefont{Lewenstein}},
  \bibinfo{journal}{Phys. Rev. Lett.} \textbf{\bibinfo{volume}{94}},
  \bibinfo{pages}{070404} (\bibinfo{year}{2005}).

\bibitem[{\citenamefont{Hazzard and
  Mueller}(2010{\natexlab{b}})}]{hazzard:techniques_quantumcriticality}
\bibinfo{author}{\bibfnamefont{K.~R.~A.} \bibnamefont{Hazzard}}
  \bibnamefont{and} \bibinfo{author}{\bibfnamefont{E.~J.}
  \bibnamefont{Mueller}}, \bibinfo{journal}{Phys. Rev. A, in press}
  (\bibinfo{year}{2010}{\natexlab{b}}).

\bibitem[{\citenamefont{Zhou and Ho}(2010)}]{zhou:signature_2010}
\bibinfo{author}{\bibfnamefont{Q.}~\bibnamefont{Zhou}} \bibnamefont{and}
  \bibinfo{author}{\bibfnamefont{T.-L.} \bibnamefont{Ho}},
  \bibinfo{journal}{Phys. Rev. Lett.} \textbf{\bibinfo{volume}{105}},
  \bibinfo{pages}{245702} (\bibinfo{year}{2010}).

\bibitem[{\citenamefont{Lercher et~al.}(2011)\citenamefont{Lercher, Takekoshi,
  Debatin, Schuster, Rameshan, Ferlaino, Grimm, and
  N{\"a}gerl}}]{lercher:production_2011}
\bibinfo{author}{\bibfnamefont{A.~D.} \bibnamefont{Lercher}},
  \bibinfo{author}{\bibfnamefont{T.}~\bibnamefont{Takekoshi}},
  \bibinfo{author}{\bibfnamefont{M.}~\bibnamefont{Debatin}},
  \bibinfo{author}{\bibfnamefont{B.}~\bibnamefont{Schuster}},
  \bibinfo{author}{\bibfnamefont{R.}~\bibnamefont{Rameshan}},
  \bibinfo{author}{\bibfnamefont{F.}~\bibnamefont{Ferlaino}},
  \bibinfo{author}{\bibfnamefont{R.}~\bibnamefont{Grimm}}, \bibnamefont{and}
  \bibinfo{author}{\bibfnamefont{H.-C.} \bibnamefont{N{\"a}gerl}}, p.
  \bibinfo{pages}{arXiv:1101.1409} (\bibinfo{year}{2011}).

\bibitem[{\citenamefont{Greif et~al.}(2011)\citenamefont{Greif, Tarruell,
  Uehlinger, J\"ordens, and Esslinger}}]{greif:probing_2011}
\bibinfo{author}{\bibfnamefont{D.}~\bibnamefont{Greif}},
  \bibinfo{author}{\bibfnamefont{L.}~\bibnamefont{Tarruell}},
  \bibinfo{author}{\bibfnamefont{T.}~\bibnamefont{Uehlinger}},
  \bibinfo{author}{\bibfnamefont{R.}~\bibnamefont{J\"ordens}},
  \bibnamefont{and}
  \bibinfo{author}{\bibfnamefont{T.}~\bibnamefont{Esslinger}},
  \bibinfo{journal}{Phys. Rev. Lett.} \textbf{\bibinfo{volume}{106}},
  \bibinfo{pages}{145302} (\bibinfo{year}{2011}).

\bibitem[{\citenamefont{Ludlow}()}]{ludlow:pc}
\bibinfo{author}{\bibfnamefont{A.}~\bibnamefont{Ludlow}},
  \bibinfo{note}{(privat comm.)}.

\bibitem[{\citenamefont{Rey et~al.}(2009)\citenamefont{Rey, Gorshkov, and
  Rubbo}}]{rey:many-body_2009}
\bibinfo{author}{\bibfnamefont{A.~M.} \bibnamefont{Rey}},
  \bibinfo{author}{\bibfnamefont{A.~V.} \bibnamefont{Gorshkov}},
  \bibnamefont{and} \bibinfo{author}{\bibfnamefont{C.}~\bibnamefont{Rubbo}},
  \bibinfo{journal}{Phys. Rev. Lett.} \textbf{\bibinfo{volume}{103}},
  \bibinfo{pages}{260402} (\bibinfo{year}{2009}).

\bibitem[{\citenamefont{Yu and Pethick}(2010)}]{yu:clock_2010}
\bibinfo{author}{\bibfnamefont{Z.}~\bibnamefont{Yu}} \bibnamefont{and}
  \bibinfo{author}{\bibfnamefont{C.~J.} \bibnamefont{Pethick}},
  \bibinfo{journal}{Phys. Rev. Lett.} \textbf{\bibinfo{volume}{104}},
  \bibinfo{pages}{010801} (\bibinfo{year}{2010}).

\bibitem[{\citenamefont{Gorshkov
  et~al.}(2011{\natexlab{a}})\citenamefont{Gorshkov, Manmana, Chen, Ye, Demler,
  Lukin, and Rey}}]{gorshkov:dipoles-tjvw-short}
\bibinfo{author}{\bibfnamefont{A.~V.} \bibnamefont{Gorshkov}},
  \bibinfo{author}{\bibfnamefont{S.~R.} \bibnamefont{Manmana}},
  \bibinfo{author}{\bibfnamefont{G.}~\bibnamefont{Chen}},
  \bibinfo{author}{\bibfnamefont{J.}~\bibnamefont{Ye}},
  \bibinfo{author}{\bibfnamefont{E.}~\bibnamefont{Demler}},
  \bibinfo{author}{\bibfnamefont{M.~D.} \bibnamefont{Lukin}}, \bibnamefont{and}
  \bibinfo{author}{\bibfnamefont{A.~M.} \bibnamefont{Rey}},
  \bibinfo{journal}{in preparation}  (\bibinfo{year}{2011}{\natexlab{a}}).

\bibitem[{\citenamefont{Gorshkov
  et~al.}(2011{\natexlab{b}})\citenamefont{Gorshkov, Manmana, Chen, Demler,
  Lukin, and Rey}}]{gorshkov:dipoles-tjvw-long}
\bibinfo{author}{\bibfnamefont{A.~V.} \bibnamefont{Gorshkov}},
  \bibinfo{author}{\bibfnamefont{S.~R.} \bibnamefont{Manmana}},
  \bibinfo{author}{\bibfnamefont{G.}~\bibnamefont{Chen}},
  \bibinfo{author}{\bibfnamefont{E.}~\bibnamefont{Demler}},
  \bibinfo{author}{\bibfnamefont{M.~D.} \bibnamefont{Lukin}}, \bibnamefont{and}
  \bibinfo{author}{\bibfnamefont{A.~M.} \bibnamefont{Rey}},
  \bibinfo{journal}{in preparation}  (\bibinfo{year}{2011}{\natexlab{b}}).

\bibitem[{\citenamefont{Idziaszek}(2009)}]{idziaszek:analytical_2009}
\bibinfo{author}{\bibfnamefont{Z.}~\bibnamefont{Idziaszek}},
  \bibinfo{journal}{Phys. Rev. A} \textbf{\bibinfo{volume}{79}},
  \bibinfo{pages}{062701} (\bibinfo{year}{2009}).

\bibitem[{\citenamefont{Lemke et~al.}(2011)\citenamefont{Lemke, Ludlow, von
  Stecher, Sherman, Rey, and Oates}}]{lemke:p-wave_2011}
\bibinfo{author}{\bibfnamefont{N.~D.} \bibnamefont{Lemke}},
  \bibinfo{author}{\bibfnamefont{A.~D.} \bibnamefont{Ludlow}},
  \bibinfo{author}{\bibfnamefont{J.}~\bibnamefont{von Stecher}},
  \bibinfo{author}{\bibfnamefont{J.~A.} \bibnamefont{Sherman}},
  \bibinfo{author}{\bibfnamefont{A.~M.} \bibnamefont{Rey}}, \bibnamefont{and}
  \bibinfo{author}{\bibfnamefont{C.~W.} \bibnamefont{Oates}},
  \bibinfo{journal}{arxiv:1105.2014}  (\bibinfo{year}{2011}).

\bibitem[{\citenamefont{Oktel et~al.}(2002)\citenamefont{Oktel, Killian,
  Kleppner, and Levitov}}]{oktel:cs-ref2}
\bibinfo{author}{\bibfnamefont{M.~O.} \bibnamefont{Oktel}},
  \bibinfo{author}{\bibfnamefont{T.~C.} \bibnamefont{Killian}},
  \bibinfo{author}{\bibfnamefont{D.}~\bibnamefont{Kleppner}}, \bibnamefont{and}
  \bibinfo{author}{\bibfnamefont{L.~S.} \bibnamefont{Levitov}},
  \bibinfo{journal}{Phys. Rev. A} \textbf{\bibinfo{volume}{65}},
  \bibinfo{pages}{033617} (\bibinfo{year}{2002}).

\bibitem[{\citenamefont{Oktel and Levitov}(1999)}]{oktel_optical_1999}
\bibinfo{author}{\bibfnamefont{M.}~\bibnamefont{Oktel}} \bibnamefont{and}
  \bibinfo{author}{\bibfnamefont{L.~S.} \bibnamefont{Levitov}},
  \bibinfo{journal}{{Phys. Rev. Lett.}} \textbf{\bibinfo{volume}{83}},
  \bibinfo{pages}{6} (\bibinfo{year}{1999}).

\bibitem[{\citenamefont{Pethick and Smith}(2001)}]{pethick:p-s-book}
\bibinfo{author}{\bibfnamefont{C.~J.} \bibnamefont{Pethick}} \bibnamefont{and}
  \bibinfo{author}{\bibfnamefont{H.}~\bibnamefont{Smith}},
  \emph{\bibinfo{title}{Bose-Einstein condensation in dilute gases}}
  (\bibinfo{publisher}{Cambridge University Press}, \bibinfo{address}{The
  Edinburgh Building, Cambridge CB2 2RU, UK}, \bibinfo{year}{2001}).

\bibitem[{\citenamefont{Mueller}(2008)}]{mueller:generic_2008}
\bibinfo{author}{\bibfnamefont{E.~J.} \bibnamefont{Mueller}},
  \bibinfo{journal}{Phys. Rev. A} \textbf{\bibinfo{volume}{78}},
  \bibinfo{pages}{045601} (\bibinfo{year}{2008}).

\bibitem[{\citenamefont{Kotochigova and
  DeMille}(2010)}]{kotochigova:electric_2010}
\bibinfo{author}{\bibfnamefont{S.}~\bibnamefont{Kotochigova}} \bibnamefont{and}
  \bibinfo{author}{\bibfnamefont{D.}~\bibnamefont{DeMille}},
  \bibinfo{journal}{Phys. Rev. A} \textbf{\bibinfo{volume}{82}},
  \bibinfo{pages}{063421} (\bibinfo{year}{2010}).

\bibitem[{\citenamefont{Kadanoff and Baym}(1962)}]{kadanoff:quantum_1962}
\bibinfo{author}{\bibfnamefont{L.~P.} \bibnamefont{Kadanoff}} \bibnamefont{and}
  \bibinfo{author}{\bibfnamefont{G.}~\bibnamefont{Baym}},
  \emph{\bibinfo{title}{Quantum statistical mechanics}} (\bibinfo{publisher}{W.
  A. Benjamin, Inc.}, \bibinfo{address}{New York}, \bibinfo{year}{1962}).

\bibitem[{\citenamefont{Pethick and Stoof}(2001)}]{pethick:pseudopot-breakdown}
\bibinfo{author}{\bibfnamefont{C.~J.} \bibnamefont{Pethick}} \bibnamefont{and}
  \bibinfo{author}{\bibfnamefont{H.~T.~C.} \bibnamefont{Stoof}},
  \bibinfo{journal}{PRA} \textbf{\bibinfo{volume}{64}}, \bibinfo{pages}{013618}
  (\bibinfo{year}{2001}).

\bibitem[{\citenamefont{Ye}()}]{ye:pc}
\bibinfo{author}{\bibfnamefont{J.}~\bibnamefont{Ye}}, \bibinfo{note}{(private
  comm.)}.

\bibitem[{\citenamefont{Greiner et~al.}(2002)\citenamefont{Greiner, Mandel,
  H{\"a}nsch, and Bloch}}]{greiner:collapse_2002}
\bibinfo{author}{\bibfnamefont{M.}~\bibnamefont{Greiner}},
  \bibinfo{author}{\bibfnamefont{O.}~\bibnamefont{Mandel}},
  \bibinfo{author}{\bibfnamefont{T.~W.} \bibnamefont{H{\"a}nsch}},
  \bibnamefont{and} \bibinfo{author}{\bibfnamefont{I.}~\bibnamefont{Bloch}},
  \bibinfo{journal}{Nature} \textbf{\bibinfo{volume}{419}}, \bibinfo{pages}{51}
  (\bibinfo{year}{2002}).

\bibitem[{\citenamefont{Sebby-Strabley
  et~al.}(2007)\citenamefont{Sebby-Strabley, Brown, Anderlini, Lee, Phillips,
  Porto, and Johnson}}]{sebby-strabley:preparing_2007}
\bibinfo{author}{\bibfnamefont{J.}~\bibnamefont{Sebby-Strabley}},
  \bibinfo{author}{\bibfnamefont{B.~L.} \bibnamefont{Brown}},
  \bibinfo{author}{\bibfnamefont{M.}~\bibnamefont{Anderlini}},
  \bibinfo{author}{\bibfnamefont{P.~J.} \bibnamefont{Lee}},
  \bibinfo{author}{\bibfnamefont{W.~D.} \bibnamefont{Phillips}},
  \bibinfo{author}{\bibfnamefont{J.~V.} \bibnamefont{Porto}}, \bibnamefont{and}
  \bibinfo{author}{\bibfnamefont{P.~R.} \bibnamefont{Johnson}},
  \bibinfo{journal}{Phys. Rev. Lett.} \textbf{\bibinfo{volume}{98}},
  \bibinfo{pages}{200405} (\bibinfo{year}{2007}).

\bibitem[{\citenamefont{Chang et~al.}(2004)\citenamefont{Chang, Ye, and
  Lukin}}]{chang:controlling_2004}
\bibinfo{author}{\bibfnamefont{D.~E.} \bibnamefont{Chang}},
  \bibinfo{author}{\bibfnamefont{J.}~\bibnamefont{Ye}}, \bibnamefont{and}
  \bibinfo{author}{\bibfnamefont{M.~D.} \bibnamefont{Lukin}},
  \bibinfo{journal}{Phys. Rev. A} \textbf{\bibinfo{volume}{69}},
  \bibinfo{pages}{023810} (\bibinfo{year}{2004}).

\end{thebibliography}

\end{document}